\documentclass[a4paper,fleqn]{cas-dc}
\usepackage[authoryear]{natbib}

\usepackage[switch]{lineno}
\usepackage{amsmath}
\usepackage{empheq}
\usepackage[section]{placeins} 
\usepackage[title]{appendix}
\usepackage{balance}
\usepackage{mathtools,microtype}
\allowdisplaybreaks
\newcommand{\Fbp}{F_{\mathrm{bp}}}
\newcommand{\Rm}{\mathcal{R}^{-}}
\newcommand{\Rp}{\mathcal{R}^{+}}
\newcommand{\Rpm}{\mathcal{R}^{\pm}}

\usepackage{etoolbox}
\newif\ifinappendix
\pretocmd{\appendix}{\inappendixtrue}{}{}

\makeatletter
\renewcommand{\@seccntformat}[1]{%
  \ifstrequal{#1}{section}{%
    \ifinappendix
      \appendixname~\csname the#1\endcsname\quad
    \else
      \csname the#1\endcsname\quad
    \fi
  }{%
    \csname the#1\endcsname\quad
  }%
}
\makeatother

\usepackage{xspace}

\def\tsc#1{\csdef{#1}{\textsc{\lowercase{#1}}\xspace}}
\tsc{WGM}
\tsc{QE}
\tsc{EP}
\tsc{PMS}
\tsc{BEC}
\tsc{DE}

\ExplSyntaxOn
\RenewDocumentEnvironment { PrelimsAbstract } { O{} }
  {
   \parindent=0pt
   { \fontsize{14pt}{16pt}\selectfont #1 }\par
   \vskip12pt
   { \fontsize{12pt}{14pt}\bfseries\selectfont\casprelimstitle } \par
   \vskip6pt
   \ifnum\theblind>0\relax
     \vspace*{\the\baselineskip}
   \else
     \seq_use:Nn \g_stm_prelimsau_seq { ,\c_space_token }
   \fi
   \vskip12pt
   \par
  }
  {}
\ExplSyntaxOff

\begin{document}
\let\WriteBookmarks\relax

\shorttitle{Finite-source filtering and Doppler modulation of infragravity waves}

\shortauthors{Midouni et~al.}

\title [mode = title]{Finite-source filtering and Doppler modulation of infragravity waves generated by breakpoint forcing}

\author[1]{Hedi Midouni}

\cormark[1]

\ead{hedi.midouni@kuleuven.be}

\credit{Conceptualization, Methodology, Software, Formal analysis, Investigation, Writing – original draft, Visualization}
\affiliation[1]{
  organization={Department of Civil Engineering, Hydraulics and Geotechnics Section, KU Leuven},
  country={Belgium}}
\affiliation[2]{organization={CSIRO Ocean and Atmosphere Flagship},
    city={Crawley},
    country={Australia}}
    
\author[2]{Stephanie Contardo}
\credit{Conceptualization, Writing – review \& editing}

\author[1]{Jaak Monbaliu}
\credit{Supervision, Writing – review \& editing, Funding acquisition}

\author[1]{Tomohiro Suzuki}

\credit{Supervision, Writing – review \& editing, Funding acquisition}

\cortext[cor1]{Corresponding author}


\begin{abstract}
Infragravity waves generated in the surf zone contribute substantially to
nearshore water-level variability, with group-scale variations in wave
breaking generating free long waves through breakpoint forcing. Existing
interpretations of this mechanism rely largely on an idealised
moving-breakpoint picture in which breaking is tied to a local saturation
relation, leaving unresolved how the finite extent and evolution of a
realistic breaking region control the radiated waves.

We first derive a local decomposition of the wave-induced forcing into breakpoint and
bound wave forcing, separating the two contributions within the same evolving
short-wave field. We then describe the breakpoint-forcing field as a finite,
evolving source through its integrated strength, centroid, cross-shore width,
and shape. This representation shows that finite source extent produces
spatial interference, as phase differences between contributions emitted from
different positions attenuate the high-frequency part of the infragravity
response and shift this attenuation toward lower frequencies as the source
widens. Source migration produces a directional Doppler shift by modifying
the arrival times of successive emissions, with shoreward motion compressing
arrivals at a shoreward observer and stretching them at a seaward observer.
Breaking depth controls both the conversion of this forcing into surface
elevation and the celerities governing these propagation-time effects. Controlled numerical experiments across a range of plane-beach slopes
reproduce the predicted signatures: broader source footprints strengthen
upper-band attenuation on both branches. The diagnosed migration history, with breaking strongest during shoreward motion, partly offsets that attenuation shoreward and reinforces it seaward.
\end{abstract}
\begin{highlights}
\item Breakpoint and bound-wave forcing are separated in the momentum balance.
\item Breaking depth controls the conversion of breakpoint forcing into surface elevation.
\item Cross-shore breaking width attenuates radiation through spatial interference.
\item Breakpoint migration compresses shoreward arrivals and stretches seaward arrivals.
\end{highlights}


\maketitle

\section{Introduction}

Infragravity waves are low-frequency surface gravity waves with periods of
approximately 20 to 250~s, whose nearshore generation is tied to the group
structure of wind-generated sea and swell \citep{Bertin2018}. Early field
observations described them as ``surf beat'' \citep{Munk1949, Tucker1950}, and
subsequent work has shown that they can dominate water-level variability in the
inner surf and swash zones \citep{Ruessink1998, Guedes2013}, modulate runup and
coastal inundation
\citep{Stockdon2006, RoeberBricker2015, Cheriton2016},
drive sediment transport and morphological change
\citep{Russell1993, Baldock2010, McCall2010}, and excite
harbour resonance \citep{Miles1974, Bowers1977, Okihiro1993, Rabinovich2009}.
These impacts make it important to understand how wave groups force long waves
and how the resulting response evolves as it propagates through the nearshore.

As short-wave groups propagate into intermediate and shallow water,
quadratic difference interactions among their components produce long-wave
motions at group frequencies \citep{Hasselmann1962}. Over a flat bed, the
corresponding equilibrium response travels with the wave group and remains
out of phase with its short-wave envelope
\citep{LonguetHiggins1962}. As the group shoals over variable depth, this
equilibrium response continuously adjusts to the changing water depth, and the total long-wave signal progressively lags behind the short-wave envelope
and grows more slowly than the local equilibrium response
\citep{MeiBenmoussa1984, Janssen2003, Battjes2004, vanDongeren2007, Zou2011,
Guerin2019, Zhang2020, Liao2021}. An equivalent description separates the
shoaling signal into the local equilibrium bound wave and the free long waves
generated as the local bound response adjusts to changes in depth
\citep{NielsenBaldock2010, Nielsen2017, Contardo2021}. In the stepwise
construction of \citet{Contardo2021}, each depth transition changes the local
bound response and generates free-wave contributions required by mass and
momentum continuity across the transition. Their superposition with the local
equilibrium response gives the same total long-wave signal commonly described
as the shoaling bound wave. \citet{Liao2023} formulated this process
continuously using a Green-function approach, treating the distributed
group-scale forcing as local sources that radiate free long waves in both
directions. As shoaling continues into shallower water and the group velocity
approaches the free long-wave celerity, the response approaches resonance and
the separation between bound and free contributions progressively loses its
meaning \citep{Liao2023}.

When the wave group reaches the breaking region, short-wave dissipation
weakens the radiation-stress variations that sustain the bound long-wave
response, allowing the long wave to decouple from the group through the
process commonly described as bound-wave release. \citet{Baldock2012}
questioned whether short-wave breaking itself triggers this release, pointing
to the strong reduction of the forced long wave after breaking and linking
progressive decoupling to the approach to shallow-water conditions. More
recently, \citet{Contardo2025} isolated a release component associated
specifically with short-wave dissipation at breaking, showing that the bound
long wave can be released from the group even away from shallow-water
resonance.

Alongside bound-wave release, a second pathway for generating free long waves
arises from group-scale variations of the breaking region and is described as
breakpoint forcing, the focus of the present paper. In the classical
moving-breakpoint model of \citet{Symonds1982}, these variations are
represented through a time-dependent breakpoint on a plane beach. Short-wave amplitude within the surf zone is prescribed by a local saturation relation; outside it, the incident amplitude varies at the group frequency. Breakpoint motion consequently changes the cross-shore extent of
the surf zone and the spatial distribution of the radiation-stress forcing
through the group cycle. The resulting free waves are generated at the group
frequency and its harmonics and radiate in both cross-shore directions,
producing a standing response shoreward of the breakpoint excursion and an
outgoing progressive wave seaward. The forced long wave associated with the
incident groups was omitted, so the breakpoint-generated free-wave response
was considered separately.

\citet{Schaeffer1993} subsequently included short-wave modulation through the
surf zone together with the incident bound long wave and showed that, for
bichromatic groups, the incident bound-wave and breakpoint-related
contributions could have similar amplitudes and partly cancel. Laboratory
experiments with bichromatic groups produced responses consistent with
generation by a time-varying breakpoint \citep{Baldock2000}, and random-wave
experiments showed a strong dependence of the outgoing infragravity response
on normalised surf-zone width \citep{BaldockHuntley2002}. Transient
focused-group experiments then directly resolved a seaward-propagating
long-wave trough emitted from the initial breaking region
\citep{Baldock2006}. More recently, \citet{Liao2023} combined their
Green-function solution with the classical moving-breakpoint forcing model,
allowing the shoreward- and seaward-propagating components to be resolved
separately. They showed that the locally generated shoreward component can
interfere constructively or destructively with the contribution arriving from
the shoaling region across the breakpoint excursion.

Experiments and models show a transition from stronger bound-wave
contributions at low normalised bed slope to stronger breakpoint-forcing
contributions at high normalised bed slope, with incident wave conditions also
affecting their relative importance
\citep{Battjes2004, Dong2009, ContardoSymonds2013}. Coral fore reefs provide a
useful setting for examining this slope dependence. At Ningaloo Reef,
\citet{Pomeroy2012} found breakpoint forcing to dominate the generation of free
infragravity waves over the fore reef relative to the shoaling bound-wave
contribution, and \citet{Masselink2019} found a similar shift toward stronger
breakpoint forcing over steeper fore-reef profiles. \citet{Liu2023} showed
that this dependence extends beyond the local breaking slope: breakpoint
forcing remained dominant when short waves broke mainly over a horizontal
reef flat after transforming across a steep fore reef. They interpreted this
persistence as evidence that the generation regime depends on the preceding
cross-shore evolution of the short-wave field.

Determining the contribution of each generation mechanism directly from the
infragravity surface elevation is difficult because free waves generated
during shoaling, released at breaking, and radiated by breakpoint forcing
occupy the same frequency band and continue to evolve together through
subsequent shoaling, partial reflection over variable bathymetry, shoreline
reflection, dissipation, and nonlinear energy transfers
\citep{Contardo2023, vanDongeren2007, Pomeroy2012, deBakker2016IG,
Rijnsdorp2022}. Consequently, field and laboratory measurements and
phase-resolving simulations contain an infragravity response in which
generation and subsequent transformation are combined, so the relative importance of the mechanisms is generally inferred
indirectly from phase relations or correlations between the short-wave
envelope and the infragravity response, which nonetheless remain informative.

Group-resolving models
\citep{vanDongeren2003, Roelvink2009, Olabarrieta2023, Reyns2023,
Marchesiello2026} provide a more direct framework, with the short-wave
envelope, long-wave hydrodynamics, and associated wave-induced forcing
represented as distinct model quantities. Several studies using
XBeach-Surfbeat have exploited this structure by restricting the
radiation-stress forcing to the active breaking region or to its exterior and
interpreting the corresponding responses as dominated by breakpoint or
bound-wave forcing, respectively
\citep{Pomeroy2012, Bertin2016, Bertin2020, Matsuba2021}. Because these masks
act on the complete wave-induced forcing and bound-wave forcing can also
contribute within the breaking region, the comparison yields an approximate separation of the two mechanisms and
remains useful for identifying the dominant generation regime, particularly
when breaking is concentrated within a narrow region.

\citet{Contardo2025} used a deliberately controlled one-dimensional linear
model with bichromatic groups, complete loss of groupiness at breaking, a
fixed breakpoint, and prescribed forcing configurations that activated the
two mechanisms separately. This setup isolated breakpoint-forced waves from a
combined release component consisting of the free waves generated during
shoaling and the bound wave released at breaking. With the breakpoint fixed,
breakpoint-forcing efficiency was independent of bed slope, group frequency,
and short-wave period; bound-wave-release efficiency decreased toward steep
normalised-slope conditions. The relative importance of the two mechanisms
was therefore governed by variations in bound-wave-release efficiency, with
phase differences between their free-wave contributions reducing the combined
response.

These two lines of work therefore motivate a local separation of breakpoint
and bound-wave forcing derived directly from the governing equations and
applicable within an evolving irregular wave field. Within the vortex-force
formalism, \citet{Midouni2025CD} proposed an initial equation-based proxy by
associating the explicit wave-momentum term proportional to breaking
dissipation with breakpoint forcing. Here, Section~\ref{sec:rs-split-short} derives such a local decomposition of
the radiation-stress forcing, and Section~\ref{subsec:link_vortex_force} relates it to that earlier formulation.

With breakpoint forcing isolated, the remainder of the paper examines how the
spatiotemporal organisation of the forcing field shapes the radiated
infragravity response. Section~\ref{sec:compact} represents its evolving
cross-shore distribution through its integrated strength, centroid, and width,
together with a normalised source shape. Section~\ref{sec:green_filters}
combines this compact representation with a Green-function formulation to
derive the radiated response and identify the roles of breaking depth,
finite-source interference, and cross-shore motion of the forcing region.
Finally, Section~\ref{sec:xbeach_mechanisms} tests these predictions with
controlled group-resolving numerical experiments across a range of
plane-beach slopes.

\section{Separating breakpoint and bound wave forcing}
\label{sec:rs-split-short}

We first develop the forcing separation in a reduced wave--flow system chosen
to expose its structure and physical interpretation. Its extension to the full
nonlinear coupled system is given in Appendix~\ref{app:full_1d_system}.

\subsection{Reduced wave--flow equations}

We consider one-dimensional cross-shore long-wave dynamics over a sloping
beach, with alongshore uniformity and normal incidence. Let $h(x)$ denote the
still-water depth, $\eta(x,t)$ the free-surface displacement, $d=h+\eta$ the
total depth, and $q(x,t)$ the total cross-shore discharge. With $g$ the
gravitational acceleration, $\rho$ the water density, and $S_{xx}$ the
cross-shore radiation stress, the reduced depth-integrated equations are
\citep{LonguetHiggins1962}
\begin{equation}
\left\{
\begin{array}{l}
\displaystyle
\frac{\partial q}{\partial t}
+ g\,h\,\frac{\partial \eta}{\partial x}
= -\,\frac{1}{\rho}\,\frac{\partial S_{xx}}{\partial x}
\\[1.2ex]
\displaystyle
\frac{\partial \eta}{\partial t}
+ \frac{\partial q}{\partial x}
= 0
\end{array}
\right.
\label{eq:rs-mom-cont-short}
\end{equation}

The short-wave field is represented by a narrow-band, unidirectional wave
group with a fixed carrier frequency. We neglect feedback from the long-wave
motion on the short waves, so that the short-wave phase and group velocities
are evaluated at the still-water depth $h$ and depend only on cross-shore
position. With depth-induced breaking as the only source of energy loss, the
short-wave energy balance is
\begin{equation}
\frac{\partial E}{\partial t}
+ \frac{\partial (c_g E)}{\partial x}
= -\,D_b
\label{eq:sw-energy}
\end{equation}
where $E$ is the short-wave energy density and $D_b$ the bulk dissipation
rate due to depth-induced breaking. For the fixed carrier frequency, the local
phase and group velocities, $c$ and $c_g$, are obtained from the linear
dispersion relation \eqref{eq:full_dispersion} using the still-water depth
$h$. To second order in wave steepness, the cross-shore radiation stress is
\begin{equation}
S_{xx} = \alpha E,
\qquad
\alpha = 2\frac{c_g}{c} - \frac{1}{2}
\label{eq:Sxx-def}
\end{equation}

\subsection{Forcing separation}
\label{subsec:split}

The separation follows by expressing the radiation-stress gradient using the
short-wave energy balance. Since $c$, $c_g$, and $\alpha$ depend on $x$
through $h(x)$ only, differentiating $S_{xx}=\alpha E$ and using
\eqref{eq:sw-energy} to eliminate $\partial E/\partial x$ gives
\begin{equation}
\begin{split}
\frac{\partial S_{xx}}{\partial x}
&= -\,\frac{\alpha\,D_b}{c_g}
- \frac{1}{c_g}\,\frac{\partial S_{xx}}{\partial t}
+ S_{xx}\,\frac{\partial}{\partial x}
\ln\!\left(\frac{\alpha}{c_g}\right)
\end{split}
\label{eq:inverse-compact}
\end{equation}

Substituting \eqref{eq:inverse-compact} into the momentum equation gives
\begin{equation}
\begin{aligned}
\frac{\partial q}{\partial t}
+ g\,h\,\frac{\partial \eta}{\partial x}
&=
F_{\mathrm{bp}} + F_{\mathrm{bw}}
\end{aligned}
\label{eq:rs-mom-split}
\end{equation}
with
\begin{equation}
\begin{aligned}
F_{\mathrm{bp}}
&\equiv
\frac{\alpha}{\rho\,c_g}\,D_b
\\[1.2ex]
F_{\mathrm{bw}}
&\equiv
\frac{1}{\rho\,c_g}
\frac{\partial S_{xx}}{\partial t}
-\frac{S_{xx}}{\rho}
\frac{\partial}{\partial x}
\ln\!\left(\frac{\alpha}{c_g}\right)
\end{aligned}
\label{eq:forcing_split_defs}
\end{equation}

The direct proportionality of \(F_{\mathrm{bp}}\) to the local breaking
dissipation \(D_b\) suggests its interpretation as the breakpoint-forcing
contribution. We therefore propose this identification, with
\(F_{\mathrm{bw}}\) collecting the remaining group-scale contributions arising
from the time evolution of the radiation stress and spatial variations of the
wave coefficients. The next subsection tests this interpretation by deriving
the free-wave response generated by each term in constant depth.

\subsection{Physical interpretation in constant depth}
\label{subsec:interpretation}

A horizontal bed removes the effects of depth variations and isolates breaking-induced free-wave generation in the same configuration used by
\citet{Contardo2025} to study bound-wave release.

On a flat bottom, \(c\), \(c_g\), and the long-wave celerity
\(c_L=\sqrt{gh}\) are constant. Combining the momentum and continuity
equations in \eqref{eq:rs-mom-cont-short} to eliminate \(q\) yields
\begin{equation}
\left(
\frac{\partial^2}{\partial t^2}
-
c_L^2\frac{\partial^2}{\partial x^2}
\right)\eta
=
\frac{1}{\rho}
\frac{\partial^2 S_{xx}}{\partial x^2}
\label{eq:wave_eq_flat}
\end{equation}

To distinguish group-bound and freely propagating motions, we follow the
decomposition used by \citet{Contardo2021}. In the absence of breaking,
\(S_{xx}=S_{xx}(x-c_g t)\), and the corresponding flat-bottom bound-wave
solution is
\begin{equation}
\eta_{\mathrm{bound}}
=
-\frac{S_{xx}}{\rho(c_L^2-c_g^2)}
\label{eq:flat_bound}
\end{equation}
which is an exact solution of \eqref{eq:wave_eq_flat}. When breaking is
introduced, we use this relation as the reference forced response and define
\(\eta_{\mathrm{free}}=\eta-\eta_{\mathrm{bound}}\). Using
\eqref{eq:sw-energy} to evaluate the derivatives of \(S_{xx}\) gives
\begin{equation}
\left(
\partial_t^2-c_L^2\partial_x^2
\right)
\eta_{\mathrm{free}}
=
-\frac{\alpha}
{\rho(c_L^2-c_g^2)}
\left(
\partial_t-c_g\partial_x
\right)D_b
\label{eq:qres}
\end{equation}

We decompose the free response as
\(\eta_{\mathrm{free}}=\eta_{\mathrm{bp}}+\eta_{\mathrm{rel}}\), where
\(\eta_{\mathrm{bp}}\) is the response to \(F_{\mathrm{bp}}\) and
\(\eta_{\mathrm{rel}}\) is the released bound-wave component remaining after
the forced response has been separated. Since
\(F_{\mathrm{bp}}=\alpha D_b/(\rho c_g)\) in constant depth,
Eq.~\eqref{eq:qres} and the breakpoint-forced equation give
\begin{equation}
\begin{aligned}
\left(
\partial_t^2-c_L^2\partial_x^2
\right)\eta_{\mathrm{bp}}
&=
-\frac{\alpha}{\rho c_g}\partial_xD_b
\\
\left(
\partial_t^2-c_L^2\partial_x^2
\right)\eta_{\mathrm{rel}}
&=
-\frac{\alpha}
{\rho(c_L^2-c_g^2)}
\left(
\partial_t-\frac{c_L^2}{c_g}\partial_x
\right)D_b
\end{aligned}
\label{eq:flat_components}
\end{equation}

For breaking dissipation with finite cross-shore support, characteristic
integration outside the breaking region with
\(\xi_{\pm}=t\mp x/c_L\), where \(+\) and \(-\) denote shoreward and seaward
propagation, gives
\begin{equation}
\begin{aligned}
\mathcal{I}_{\pm}(\xi_{\pm})
&=
\int
D_b\left(
y,\xi_{\pm}\pm\frac{y}{c_L}
\right)\,dy
\\
\eta_{\pm}^{\mathrm{bp}}
&=
\pm\frac{\alpha}{2\rho c_g c_L^2}
\mathcal{I}_{\pm}
\\
\eta_{\pm}^{\mathrm{rel}}
&=
\mp\frac{\alpha}
{2\rho c_g c_L(c_L\mp c_g)}
\mathcal{I}_{\pm}
\end{aligned}
\label{eq:flat_branches}
\end{equation}

On each propagation branch, the breakpoint-forced and released components
therefore have the same waveform and are in antiphase, with
\begin{equation}
\frac{\eta_{\pm}^{\mathrm{rel}}}
{\eta_{\pm}^{\mathrm{bp}}}
=
-\frac{c_L}{c_L\mp c_g}
\label{eq:ratio_green}
\end{equation}
The release-to-breakpoint amplitude ratio is consequently larger shoreward,
where it is \(c_L/(c_L-c_g)\), than seaward, where it is
\(c_L/(c_L+c_g)\).

For a breaking source that is symmetric about a fixed cross-shore position,
\(\mathcal{I}_{+}\) and \(\mathcal{I}_{-}\) have the same waveform apart from
a propagation-time shift. The breakpoint-forced response then has equal
amplitude in the two directions, whereas the released bound-wave response is
asymmetric, with
\begin{equation}
\frac{|\eta_-^{\mathrm{rel}}|}
{|\eta_+^{\mathrm{rel}}|}
=
\frac{c_L-c_g}{c_L+c_g}
\label{eq:flat_release_ratio}
\end{equation}
which agrees with the flat-bed release ratio of
\citet[][Eq.~B7]{Contardo2025}. The released bound wave is therefore stronger
shoreward than seaward.

The constant-depth solution thus gives the expected signatures of the two
breaking-related contributions: symmetric bidirectional breakpoint-forced
radiation, a stronger shoreward released bound wave, and antiphase between the
two components on each branch. Together with the agreement with the established
flat-bed release solution, these results provide a first consistency check on
the physical interpretation of the forcing separation introduced above.

\subsection{Forcing signatures on a sloping beach}
\label{subsec:forcing_signatures}

The preceding analytical exercise establishes the expected relationship
between breakpoint forcing and bound-wave release under idealised
constant-depth conditions. We next ask whether the same physical picture
remains identifiable when irregular waves shoal and break over a sloping
beach.

We address this question with the in-house group-resolving model described in
Appendix~\ref{app:full_1d_system}, first using the complete radiation-stress
forcing and comparing the response with SWASH, a well-validated
phase-resolving model run with two vertical layers \citep{Zijlema2011}. We
then use the decomposition as a physical diagnostic by repeating the
group-resolving calculation with only $F_{\mathrm{bp}}$ or
$F_{\mathrm{bw}}$ included in the long-wave momentum equation, so that the
propagating features in the complete solution can be related to each forcing
contribution.

The physical configuration consists of a $1{:}20$ planar beach with the
offshore boundary in $20$~m water depth. The incident wave conditions follow
a JONSWAP spectrum with $H_{m0}=2$~m, $T_p=10$~s, and $\gamma=3.3$.

Both models use $\Delta x=2$~m. SWASH needs this value to resolve individual
short waves; the group-resolving model evolves only the wave envelope and
would admit a coarser grid, but a common grid is kept to ease the
comparison. SWASH adopts the default breaking settings of \citet{Smit2013},
and the group-resolving model uses the \texttt{roelvink2} formulation of
XBeach-Surfbeat \citep{Roelvink2009}, calibrated against the SWASH
short-wave-height profile and surf-zone group modulation ($\gamma_b=0.4$,
$n=6$, $\alpha_b=1.1$). We include nonlinear long-wave dynamics, omit
short-wave current feedback, and use $2$~h of post-spin-up output.

The same wave realisation is used in both models by prescribing a JONSWAP spectrum at the SWASH offshore boundary, exporting the resulting free-wave components above $1/(2T_p)=0.05$~Hz, and using the Hilbert envelope of the corresponding time series to prescribe the time-varying short-wave energy at the boundary of the group-resolving model.
SWASH includes the incident equilibrium bound wave following
\citet{Vasarmidis2024}, and the full and bound-wave-only group-resolving
calculations impose the equilibrium solution given by
Eq.~\eqref{eq:flat_bound}; the breakpoint-only calculation omits this contribution.

For post-processing, the same $0.05$~Hz cutoff separates the sea-swell and
infragravity bands in all calculations. We obtain the short-wave envelope
from the high-pass-filtered surface elevation in SWASH and directly from the
group-resolving model, then construct each map from the lagged
cross-correlation between the infragravity elevation and the corresponding
offshore short-wave envelope.

\begin{figure}[pos=htbp]
    \centering
    \includegraphics[width=\linewidth]{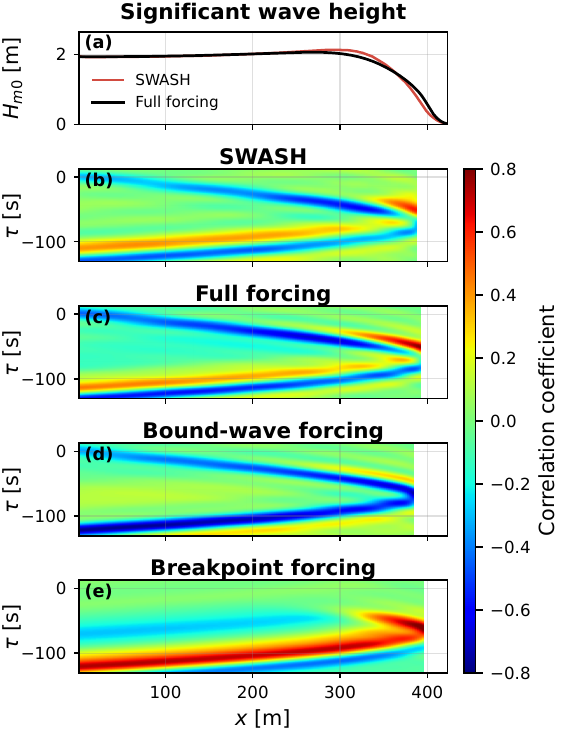}
    \caption{Comparison between SWASH and the group-resolving model, followed
    by forcing-component diagnostics on the $1{:}20$ planar beach.
    (a) Cross-shore significant wave height $H_{m0}$ in SWASH and
    the full group-resolving calculation. (b,c) Lagged cross-correlations
    between the offshore short-wave envelope and the local infragravity
    elevation in the same two calculations. (d,e) Corresponding maps with bound-wave forcing only and breakpoint forcing only, respectively.
    Each record extends to its own continuously wet support.}
    \label{fig:crosscorr_app}
\end{figure}

Figure~\ref{fig:crosscorr_app} first compares the two complete calculations, then
shows the group-resolving response to each forcing contribution separately.
Panel~(a) shows good agreement between the cross-shore
significant-wave-height profiles. Panels~(b) and (c) likewise contain the same
dominant correlation patterns, with some local differences in their amplitude
and position. To interpret these mixed signatures, we now examine the two
forcing contributions separately.

In the bound-wave calculation (d), the dominant
negative ridge follows the shoreward propagation of the bound-wave depression
and remains negatively correlated with the offshore group signal after
reflection at the shoreline. Much weaker positive ridges appear during
shoaling, with slopes and arrival times consistent with free long waves
generated as the bound response adjusts to the varying depth
\citep{Contardo2021}.

The breakpoint-forcing calculation (e) displays the bidirectional response
predicted by the constant-depth analysis, with a positively correlated branch
propagating shoreward from the breaking region and a negatively correlated
branch propagating seaward. The shoreward branch reaches the shoreline
before the dominant negative ridge in (d), consistent with the breakpoint
response propagating as a free long wave whereas the phase-lagged shoaling bound
response remains tied to the group velocity $c_g<c_L$ over most of the
shoaling path \citep{Liao2023}. As the positive branch approaches the coast, a seaward-propagating return develops progressively through partial reflection over the inner slope and continues through reflection at the shoreline. A weaker negative return appears later along a
distinct offshore trajectory. Multiple partial reflections over the varying
bathymetry offer a plausible explanation for this feature
\citep{Contardo2023}: the first positive return may generate an
opposite-polarity shoreward component while crossing the slope, which would
then reach the shoreline and reflect offshore once more. The correlation map
does not resolve the individual reflection sequence, although the timing and
polarity of the later branch are consistent with such a higher-order path.

The full response combines these bound-wave and breakpoint-forced branches,
including their partial cancellation and successive reflections. Their close
counterparts in SWASH show that the spatial organisation revealed by the
decomposition also appears in a model with a more complete representation of
nearshore wave dynamics. The decomposition therefore provides a useful
physical diagnostic, allowing the main features of the complete long wave field to
be traced back to their respective forcing mechanisms, a distinction that cannot be obtained directly from the SWASH simulation alone.

\section{Compact representation of breakpoint forcing}
\label{sec:compact}

\begin{figure*}[pos=htbp]
\centering
\includegraphics[width=\textwidth]{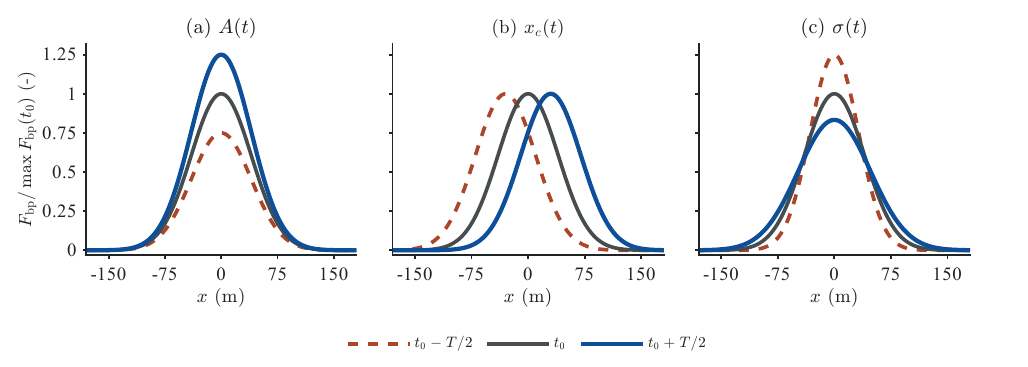}
\caption{Schematic location--scale representation of a Gaussian footprint. Each panel varies one descriptor while holding
the other two fixed: (a) \(A\) sets the integrated strength, (b) \(x_c\)
translates the footprint, and (c) \(\sigma\) changes its cross-shore width.}
\label{fig:schematic_3params}
\end{figure*}

The remainder of the analysis focuses on breakpoint forcing. Considering only
\(F_{\mathrm{bp}}\) in the momentum balance \eqref{eq:rs-mom-split} and
combining it with continuity in \eqref{eq:rs-mom-cont-short} gives
\begin{equation}
\frac{\partial^2 \eta}{\partial t^2}
-
\frac{\partial}{\partial x}
\left(
g h\,\frac{\partial \eta}{\partial x}
\right)
=
-\,\frac{\partial F_{\mathrm{bp}}(x,t)}{\partial x}
\label{eq:eta_wave}
\end{equation}

\subsection{Group-scale organisation of breaking}
\label{subsec:source_evolution}

Depth-induced breaking occurs through the rapid deformation and collapse of
individual short-wave crests as they shoal into shallow water. The associated
momentum transfer and loss of organised short-wave energy are therefore
localised and intermittent on the short-wave phase scale, with the breaking
state varying from one wave to the next. To understand how breaking drives
long-wave motions, the short-wave phase must be filtered out so that the
slower organisation of breaking can be examined on the time scale over which
the long-wave forcing evolves.

At this slower scale, breaking exhibits a systematic evolution over the
passage of a wave group. As the more energetic part of the group reaches shallow water, the larger
waves begin breaking farther offshore, so the active region extends seaward.
These waves then dissipate most strongly as they travel shoreward through the
surf zone, so the strongest breaking accompanies a shoreward shift of the
active region, which weakens and contracts after the energetic part of the
group has passed, until the next group renews breaking farther offshore. This group-scale migration and deformation emerge
from the underlying sequence of individual breaking waves and are illustrated
by the phase-resolving simulation shown in Figure~1 of \citet{Moura2018}.

In the wave-averaged description used here, \(D_b(x,t)\) captures this slower
spatiotemporal organisation of the breaking dissipation. Because
\(F_{\mathrm{bp}}\) is proportional to \(D_b\) through
\eqref{eq:forcing_split_defs}, the breakpoint-forcing field reflects both the intensity of dissipation and its cross-shore distribution over the group cycle.

\subsection{Location--scale representation of breakpoint forcing}
\label{subsec:ansatz}

The group-scale behaviour described above suggests representing the
distributed forcing field through a small number of descriptors of its
evolution and spatial structure. We call each instantaneous cross-shore
distribution of \(F_{\mathrm{bp}}\) a forcing footprint and characterise it
by an integrated strength, a cross-shore centroid, and a spatial width,
together with a normalised reference shape. Denoting the three time-dependent
descriptors by \(A(t)\), \(x_c(t)\), and \(\sigma(t)\), respectively, we write
the location--scale, or shift-and-stretch, representation \citep{Casella2002}
\begin{equation}
F_{\mathrm{bp}}(x,t)
\approx
\frac{A(t)}{\sigma(t)}
\varphi\!\left(\frac{x-x_c(t)}{\sigma(t)}\right)
\label{eq:Fbp_ansatz}
\end{equation}

Here \(\zeta=[x-x_c(t)]/\sigma(t)\) is the normalised cross-shore coordinate
and \(\varphi(\zeta)\geq0\) the corresponding dimensionless reference shape,
normalised such that \(\int\varphi(\zeta)\,d\zeta=1\).

The three time-dependent descriptors are defined directly from the
instantaneous forcing footprint through its zeroth, first, and second centred
spatial moments:
\begin{equation}
\begin{aligned}
A(t) &= \int F_{\mathrm{bp}}(x,t)\,dx \\
x_c(t) &= \frac{1}{A(t)}\int x\,F_{\mathrm{bp}}(x,t)\,dx \\
\sigma^2(t) &= \frac{1}{A(t)}
\int \bigl(x-x_c(t)\bigr)^2 F_{\mathrm{bp}}(x,t)\,dx
\end{aligned}
\label{eq:moment_defs}
\end{equation}

for active instants with \(A(t)>0\). To align the location and scale of
\(\varphi\) with these moment definitions, we impose
\(\int\zeta\varphi(\zeta)\,d\zeta=0\) and
\(\int\zeta^2\varphi(\zeta)\,d\zeta=1\). The unit-integral normalisation makes
\(A(t)\) the integrated forcing strength, and the zero-mean and
unit-variance conditions make \(x_c(t)\) the forcing centroid and
\(\sigma(t)\) its root-mean-square cross-shore width.

Figure~\ref{fig:schematic_3params} illustrates these roles for a Gaussian
reference shape by varying one descriptor at a time. Changing \(A\) scales
the forcing strength, changing \(x_c\) translates the footprint, and changing
\(\sigma\) stretches or contracts it in the cross-shore direction. Together, the three histories describe changes in strength, position, and width, and \(\varphi\) carries the normalised spatial structure of the
footprint. The location--scale approximation therefore assumes that the primary
breakpoint-forcing footprints remain approximately self-similar in the
coordinate \(\zeta\), an assumption we test below against
\(F_{\mathrm{bp}}(x,t)\) diagnosed by the group-resolving model.

\subsection{Validation against diagnosed breakpoint forcing}
\label{subsec:ansatz_validation}
\begin{figure*}[pos=htbp]
\centering
\includegraphics[width=\textwidth]{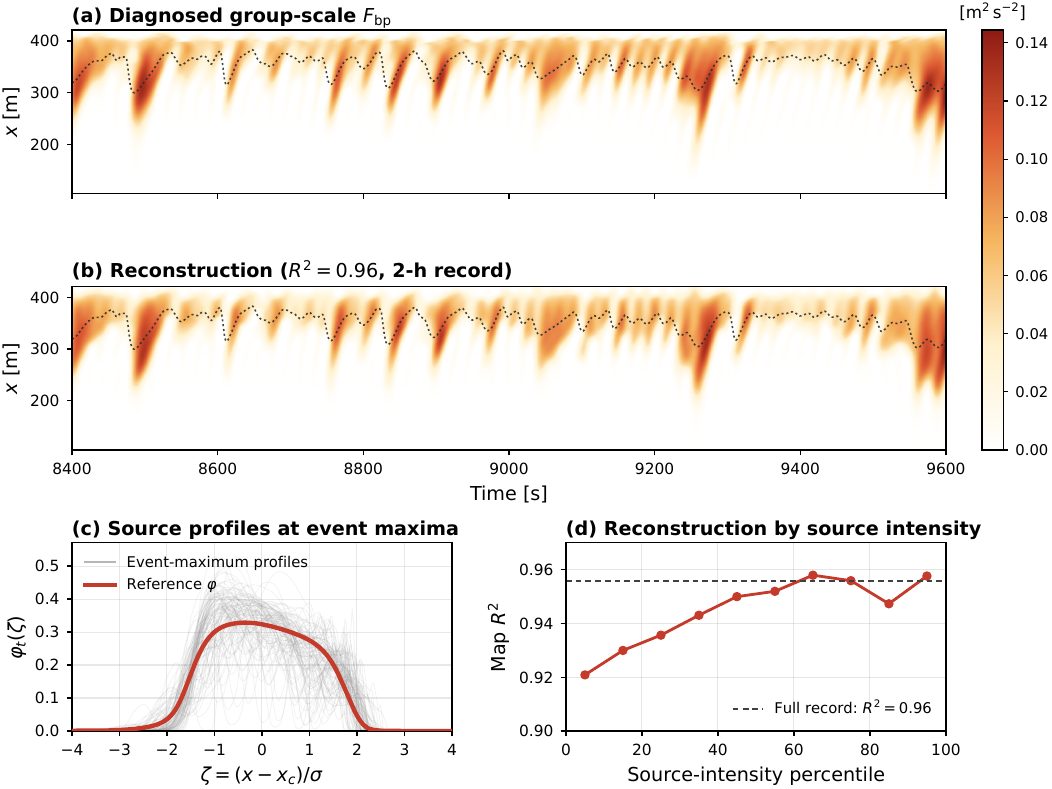}
\caption{Validation of the location--scale approximation for the \(1{:}20\)
plane-beach case introduced in Section~\ref{subsec:forcing_signatures}.
(a,b) Diagnosed and reconstructed group-scale breakpoint forcing during a
\(1200\)-s interval of strong source activity; dotted curves mark the
diagnosed centroid. The \(R^2\) reported in (b) is evaluated over the complete
\(2\)-h record. (c) Normalised profiles at the maxima of the group-scale
source events (grey), together with the full-record least-squares reference
shape \(\varphi\) (red). (d) Reconstruction \(R^2\) by source-intensity
decile over the complete record; the dashed line marks the overall value.}
\label{fig:ansatz_validation}
\end{figure*}
We assess the location--scale approximation in the \(1{:}20\) plane-beach
case introduced in Section~\ref{subsec:forcing_signatures}. The
group-resolving calculation with complete wave forcing and SWASH showed
similar nearshore correlation patterns in that case, with local differences
between the two models. The group-resolving model records
\(F_{\mathrm{bp}}(x,t)\) within the full calculation, allowing us to test
Eq.~\eqref{eq:Fbp_ansatz} directly under irregular waves.

At each output time, we use a spatially smoothed field only to identify the
connected forcing footprint around the dominant maximum, then calculate
\(A(t)\), \(x_c(t)\), and \(\sigma(t)\) from the unsmoothed values through
Eq.~\eqref{eq:moment_defs}. With these three histories fixed, we obtain a
single reference shape \(\varphi\) by minimising the space--time \(L^2\)
error over the complete unfiltered \(2\)-h record. In normalised coordinates,
the instantaneous profiles are
\begin{equation}
\varphi_t(\zeta)
=
\frac{\sigma(t)}{A(t)}
F_{\mathrm{bp}}\!\left[x_c(t)+\sigma(t)\zeta,t\right]
\label{eq:instantaneous_profile}
\end{equation}
and the minimiser is their average weighted by \(A^2(t)/\sigma(t)\).

We reconstruct the forcing on the native output grid before applying the same
\(0.05\)~Hz low-pass filter to the diagnosed and reconstructed fields.
Figure~\ref{fig:ansatz_validation} compares the two fields, with
panels~(a) and (b) showing a \(1200\)-s interval. To illustrate the
variability of the source shape without averaging over events, panel~(c)
shows one unsmoothed normalised profile at the maximum of each group-scale
source event, defined as a contiguous interval during which \(A(t)\) exceeds
its time mean. We fit \(\varphi\) using every output time.

The reconstruction in panel~(b) closely reproduces the diagnosed field in
panel~(a), even though source strength, position, and width vary strongly from
one event to the next, and the \(R^2\) reported in (b) confirms this agreement
over the complete record. The event profiles in panel~(c) remain concentrated
around \(\varphi\), and panel~(d) shows that the fit remains high across
source-intensity deciles. Three histories and a single reference shape
therefore capture the dominant space--time organisation of the diagnosed
breakpoint forcing.

\section{Spectral response to a finite, evolving breakpoint source}
\label{sec:green_filters}

The location--scale representation developed in Section~\ref{sec:compact}
reduces the evolving breakpoint-forcing field to three time histories and a
fixed reference shape. Combined with the
long-wave equation in Eq.~\eqref{eq:eta_wave}, this description allows us to
trace how the space--time organisation of the source enters the spectrum of
the radiated waves. Random wave groups distribute the forcing over a range of frequencies, for
which a frequency-domain treatment gives the most direct description. Throughout this section, the source is prescribed;
its dependence on the offshore short-wave field and on the breaking closure
lies outside the radiation problem considered here.

The linearity of Eq.~\eqref{eq:eta_wave} makes a Green function a natural way
to assemble the contributions emitted across the breaking region. We first
derive this representation and then specialise it to a horizontal bed, where
uniform propagation isolates the role of source geometry.
Section~\ref{subsec:lowpass} examines the effect of source width with a fixed
centroid, Section~\ref{subsec:doppler} restores centroid motion, and
Section~\ref{subsec:vardepth} recasts both mechanisms in travel-time
coordinates over variable bathymetry.

\subsection{Green-function representation}
\label{subsec:green}

We transform Eq.~\eqref{eq:eta_wave} in time using, for any dimensional field
\(\psi(x,t)\), the convention
\begin{equation}
\widehat{\psi}(x,\omega)
\equiv
\int_{-\infty}^{\infty}
\psi(x,t)\,e^{i\omega t}\,dt
\label{eq:ft_def_time}
\end{equation}
The ordinary frequency is \(f=\omega/(2\pi)\), and the inverse transform
uses \(e^{-i\omega t}\), so that \(\partial_t\) becomes \(-i\omega\).
The long-wave equation therefore takes the form
\begin{equation}
-\omega^2\widehat\eta
-\frac{\partial}{\partial x}\!\left(
gh\,\frac{\partial\widehat\eta}{\partial x}
\right)
=
-\frac{\partial\widehat\Fbp}{\partial x}
\label{eq:eta_freq}
\end{equation}

At each \(\omega\), Eq.~\eqref{eq:eta_freq} is a linear boundary-value
problem whose distributed response can be assembled from point-source
solutions. Following the Green-function approach used by \citet{Liao2023}
for group-induced long waves over variable bathymetry, we take \(x\) as the
observer position and \(y\) as the source position, and define
\(G(x,y;\omega)\) through
\begin{equation}
-\omega^2 G
-\frac{\partial}{\partial x}\!\left[
gh(x)\,\frac{\partial G}{\partial x}
\right]
=
\delta(x-y)
\label{eq:green_def}
\end{equation}
with outgoing radiation conditions at both ends of the domain. The
differential operator is reciprocal, and its outgoing Green function satisfies
\(G(x,y;\omega)=G(y,x;\omega)\), allowing the same equation to be read in
the source coordinate when deriving the distributed response.

Let \(\mathcal{S}\) contain the breakpoint-forcing region. Applying Green's
identity in the source coordinate and integrating the forcing derivative by
parts, with \(\widehat\Fbp=0\) on \(\partial\mathcal{S}\), gives
\begin{equation}
\widehat\eta(x,\omega)
=
\int_{\mathcal{S}}
\frac{\partial G(x,y;\omega)}{\partial y}
\,\widehat\Fbp(y,\omega)\,dy
\label{eq:green_rep}
\end{equation}
Here \(\widehat\Fbp\) weights the contribution from each source position,
and \(\partial_yG\) carries that contribution to the observer with the gain
and phase accumulated along its propagation path. The derivative of the
Green function follows from the forcing gradient in
Eq.~\eqref{eq:eta_freq}; a localised contribution consequently has a
dipole-like response, with opposite polarities radiated to either side, as in
the time-domain result of Section~\ref{subsec:interpretation}.

\subsection{Radiation in constant depth}
\label{subsec:radiation}

As in Section~\ref{subsec:interpretation}, we first consider a horizontal bed
so that the relation between the spatiotemporal structure of the
breakpoint-forcing field and the radiated waves can be examined independently
of bathymetric transformations. Although this constant-depth problem admits a
simple direct solution, we formulate it using a Green function from the outset
so that the same framework can later be extended to variable bathymetry, where
the propagation phase and gain depend on source position.

For \(h=h_c\) constant, \(c_L=\sqrt{gh_c}\) and
\(k=\omega/c_L\), the outgoing Green function for
Eq.~\eqref{eq:eta_freq} is
\begin{equation}
G(x,y;\omega)
=
\frac{i}{2k c_L^2}\,e^{ik|x-y|}
\label{eq:G_flat}
\end{equation}
For forcing confined to \([x_1,x_2]\), substitution into
Eq.~\eqref{eq:green_rep} gives
\begin{equation}
\widehat\eta(x,\omega)
=
\begin{cases}
\Rp(\omega)\,e^{ikx}, & x>x_2\\[2pt]
\Rm(\omega)\,e^{-ikx}, & x<x_1
\end{cases}
\label{eq:branches}
\end{equation}
With \(x\) increasing shoreward and time dependence \(e^{-i\omega t}\),
\(e^{ikx}\) propagates shoreward and \(e^{-ikx}\) seaward, with amplitudes
\begin{equation}
\Rpm(\omega)
=
\pm\frac{1}{2c_L^2}
\int_{x_1}^{x_2}
e^{\mp iky}\,\widehat\Fbp(y,\omega)\,dy
\label{eq:Rpm}
\end{equation}
Both amplitudes use the phase of the outgoing wave extrapolated to the common
origin \(x=0\), so translating the source rotates their phase. In the
point-source limit, Eq.~\eqref{eq:Rpm} gives two branches with equal magnitude
and opposite sign, in agreement with the time-domain result of
Section~\ref{subsec:interpretation}.

We now insert the source representation of Section~\ref{sec:compact}.
Writing the temporal transform of \(\Fbp\) explicitly and exchanging the
order of integration leaves, at each emission time, the spatial transform of
one translated and dilated reference shape. Defining
\(\widehat\varphi(q)\equiv\int_{-\infty}^{\infty}
\varphi(\zeta)e^{iq\zeta}\,d\zeta\), the shift and scaling properties of
this transform give
\begin{equation}
\int e^{\mp iky}\,\Fbp(y,t)\,dy
=
A(t)\,e^{\mp ikx_c(t)}
\,\widehat\varphi\!\big(\mp k\sigma(t)\big)
\label{eq:shiftscale}
\end{equation}
The integrated strength \(A(t)\) sets the amplitude, translation by
\(x_c(t)\) supplies the phase factor, and the width \(\sigma(t)\) rescales
the argument of \(\widehat\varphi\). The factor introduced by stretching
the cross-shore coordinate cancels the \(1/\sigma(t)\) normalisation of the
ansatz, preserving the integrated forcing for every width. Substitution into
Eq.~\eqref{eq:Rpm} then reduces the outgoing amplitudes to one integral over
the complete source history:
\begin{equation}
\begin{aligned}
\Rpm(\omega)={}&
\pm\frac{1}{2c_L^2}
\int_{-\infty}^{\infty}
A(t)\,
\underbrace{e^{\mp ikx_c(t)}}_{
\substack{\text{centroid phase}}}
\\
&\quad\times
\underbrace{\widehat\varphi\!\big(\mp k\sigma(t)\big)}_{
\substack{\text{finite-width}\\\text{coherence factor}}}
\,e^{i\omega t}\,dt
\end{aligned}
\label{eq:single}
\end{equation}

Equation~\eqref{eq:single} keeps the evolving width and centroid inside the
temporal integral, where they modulate the contributions from the strength
history \(A(t)\). In the following, we first hold the centroid fixed to isolate interference
across a finite-width footprint, then restore its motion to follow the resulting change in arrival timing.

\subsection{Interference across a finite source}
\label{subsec:lowpass}

\begin{figure*}[pos=htbp]
\centering
\includegraphics[width=\textwidth]{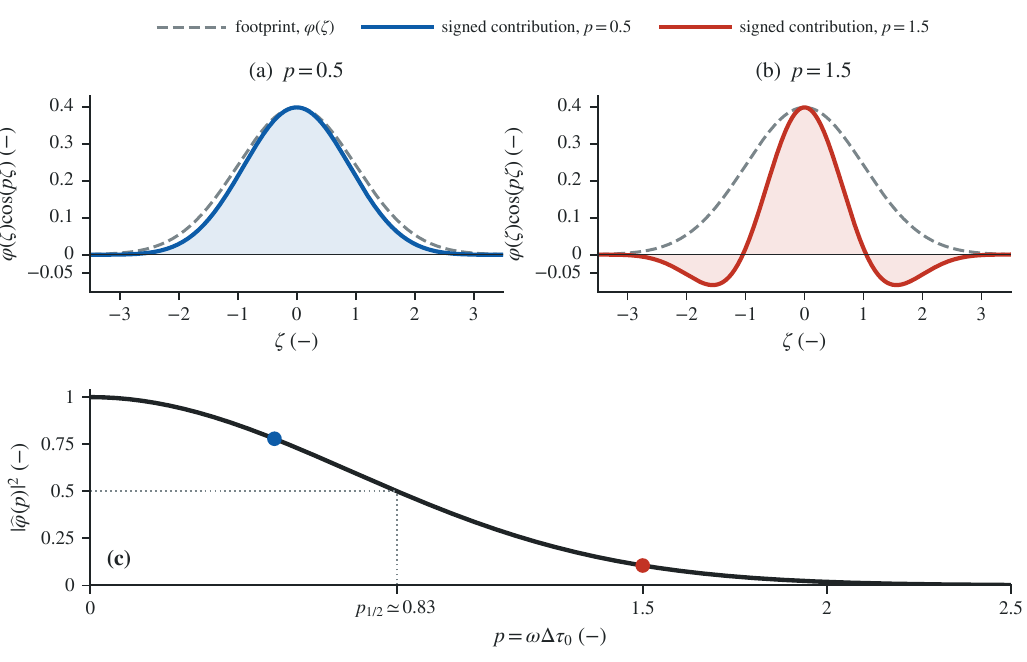}
\caption{Interference across a Gaussian breakpoint-forcing footprint.
(a,b) Signed contributions \(\varphi(\zeta)\cos(p\zeta)\) at phase
spreads \(p=k\sigma=0.5\) and \(1.5\); the dashed curve shows the source
shape. (c) Fraction of radiated power remaining after spatial interference;
dotted guides mark the first half-power point \(p_{1/2}\).}
\label{fig:filter}
\end{figure*}

The factor \(\widehat\varphi(\mp k\sigma)\) in
Eq.~\eqref{eq:single} sums the contributions emitted across the forcing
footprint. When the footprint is narrow compared with the radiated
wavelength, propagation-time differences across the source are small and the
contributions reach an observer with similar phases. As the phase variation
increases, contributions from different parts of the source begin to cancel.

For one emission time, write
\(y=x_c(t)+\sigma(t)\zeta\) and define the travel-time width of the
footprint and the corresponding phase spread at frequency \(\omega\):
\begin{equation}
\begin{aligned}
&\Delta\tau(t)\equiv\frac{\sigma(t)}{c_L}
\\
&p(t)\equiv k\sigma(t)
=\omega\Delta\tau(t)
\end{aligned}
\label{eq:dtau_inst}
\end{equation}
A contribution emitted at \(\zeta\) then carries the phase
\(\mp p\zeta\) relative to one emitted at the centroid, giving
\begin{equation}
\begin{aligned}
\widehat\varphi(\mp p)
&=
\int_{-\infty}^{\infty}
\varphi(\zeta)e^{\mp ip\zeta}\,d\zeta
\\
&=
\int_{-\infty}^{\infty}
\varphi(\zeta)\cos(p\zeta)\,d\zeta
\mp i
\int_{-\infty}^{\infty}
\varphi(\zeta)\sin(p\zeta)\,d\zeta
\end{aligned}
\label{eq:coherence_sum}
\end{equation}
Because \(\varphi(\zeta)\,d\zeta\) represents the fraction of integrated
forcing carried by each element of the footprint, this integral sums their
propagation phases with the appropriate forcing weight. The parameter \(p\)
is the phase spread over one root-mean-square source width.
Figure~\ref{fig:filter} illustrates this sum for a symmetric Gaussian
footprint, for which the sine integral vanishes and only the cosine term
remains. At \(p=0.5\), most contributions have the same sign and add
coherently; at \(p=1.5\), the negative lobes cancel a substantial part of the
total, producing the reduction in radiated power shown in panel~(c).

The flat-bed relation \(k=\omega/c_L\) maps this spatial phase spread onto
temporal frequency. For a fixed centroid and width, the centroid phase
\(e^{\mp ikx_c}\) and finite-width factor
\(\widehat\varphi(\mp k\sigma)\) can be taken outside the temporal integral in
Eq.~\eqref{eq:single}. Writing \(\widehat A\) for the temporal transform of
\(A\) gives
\begin{equation}
\begin{aligned}
&\Rpm(\omega)
=
\pm\frac{e^{\mp ikx_c}}{2c_L^2}\,
\widehat A(\omega)\,
\widehat\varphi(\mp\omega\Delta\tau)
\\
&\frac{|\Rpm(\omega)|^2}
{|\Rpm_{\mathrm{pt}}(\omega)|^2}
=
\big|\widehat\varphi(\omega\Delta\tau)\big|^2
\end{aligned}
\label{eq:fixed_source_spectrum}
\end{equation}
where \(\Rpm_{\mathrm{pt}}\) denotes the radiation from a point source with
the same strength history and centroid. The resulting power ratio measures
the fraction of radiation remaining after finite-source interference. Since
\(\varphi\) is real,
\(\lvert\widehat\varphi(-p)\rvert=\lvert\widehat\varphi(p)\rvert\), so the
power attenuation is identical on the two branches, including for an
asymmetric footprint.

In the narrow-source limit \(p\ll1\), expanding Eq.~\eqref{eq:coherence_sum}
and using the unit-integral, zero-mean, and unit-variance conditions gives
\begin{equation}
\big|\widehat\varphi(p)\big|^2
=
1-p^2+o(p^2)
\qquad (p\to0)
\label{eq:coherent_expansion}
\end{equation}
Thus, at a given frequency, the leading coherence loss for a sufficiently
narrow source is set entirely by its width and is independent of the detailed
shape \(\varphi\). The source shape enters at higher order as the phase spread
increases, controlling the subsequent attenuation and any zeros or secondary
lobes. Since \(p=2\pi f\Delta\tau\), increasing the source width shifts this
attenuation toward lower frequencies.

When the width varies in time, the coherence factor remains inside the
temporal integral in Eq.~\eqref{eq:single}, so an evolving source has no single finite-width transfer function. For comparison between sources,
we define the reference widths
\begin{equation}
\begin{aligned}
&\Delta\tau_0^2\equiv
\frac{\displaystyle\int A(t)\,\Delta\tau^2(t)\,dt}
{\displaystyle\int A(t)\,dt}
\\
&\sigma_0^2\equiv
\frac{\displaystyle\int A(t)\,\sigma^2(t)\,dt}
{\displaystyle\int A(t)\,dt}
=
c_L^2\Delta\tau_0^2
\end{aligned}
\label{eq:dtau0}
\end{equation}
where the final equality applies in constant depth.

For a given reference shape, let \(p_{1/2}>0\) be the first value satisfying
\(\lvert\widehat\varphi(p_{1/2})\rvert^2=1/2\). The corresponding reference
half-power frequency is
\begin{equation}
f_{1/2}
=
\frac{p_{1/2}}{2\pi\Delta\tau_0}
\label{eq:fhalf}
\end{equation}
For a source with variable width, \(f_{1/2}\) provides a convenient comparison
scale based on \(\Delta\tau_0\), rather than a unique cutoff of the complete
response. For the unit-variance Gaussian
\(\varphi(\zeta)=(2\pi)^{-1/2}\exp(-\zeta^2/2)\),
\(\lvert\widehat\varphi(p)\rvert^2=\exp(-p^2)\), giving
\(p_{1/2}=\sqrt{\ln2}\) and
\(f_{1/2}\approx0.1325/\Delta\tau_0\).

\begin{figure}[pos=htbp]
\centering
\includegraphics[width=\columnwidth]{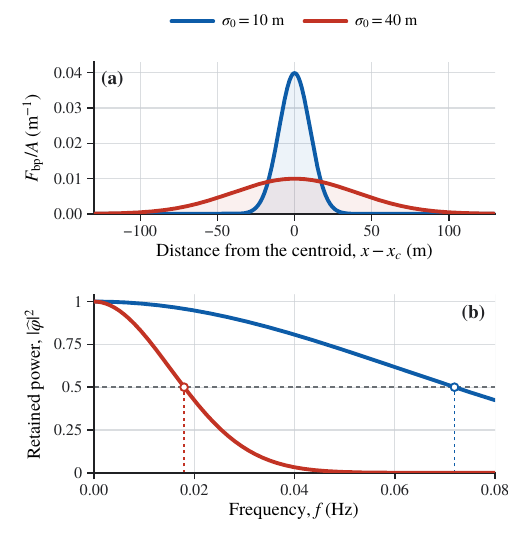}
\caption{Finite-source interference on a dimensional frequency axis.
(a) Gaussian footprints with the same integrated strength
\(A\) and centroid but \(\sigma_0=10~\mathrm{m}\) and
\(40~\mathrm{m}\). (b) Fraction of radiated power remaining at frequency
\(f\) at the constant depth \(h=3~\mathrm{m}\). The guides mark the
corresponding half-power frequencies.}
\label{fig:spatial_temporal_filter}
\end{figure}

Figure~\ref{fig:spatial_temporal_filter} places this scaling on a dimensional
frequency axis for \(h=3~\mathrm{m}\), where
\(c_L=5.42~\mathrm{m\,s^{-1}}\), and compares two Gaussian footprints with
the same integrated strength and centroid.

The fourfold increase in source width shifts the half-power frequency downward
by the same factor, illustrating directly how a broader breaking region attenuates progressively lower frequencies.

\subsection{Doppler shifts from breakpoint migration}
\label{subsec:doppler}

\begin{figure*}[pos=htbp]
\centering
\includegraphics[width=\textwidth]{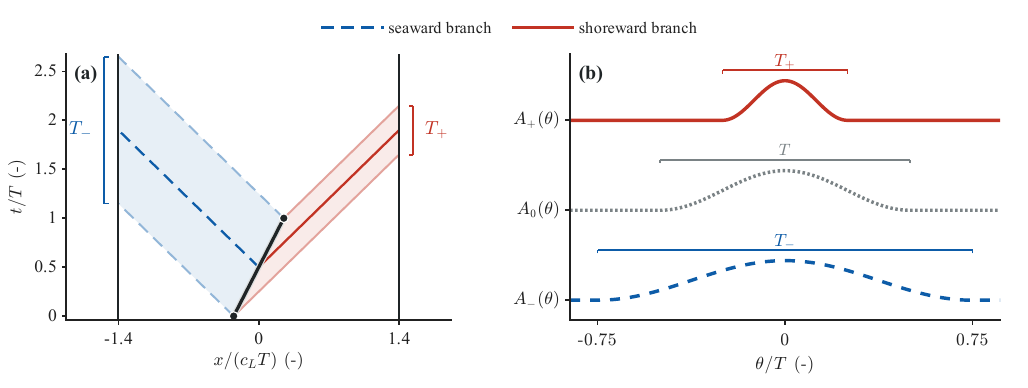}
\caption{Doppler shift produced by one active shoreward source leg with
\(v/c_L=0.5\). The characteristics in (a) give the seaward duration \(T_-\)
and shoreward duration \(T_+\), and (b) shows the same source history in the
stationary, shoreward, and seaward arrival-time coordinates; curves are
normalised and offset.}
\label{fig:doppler}
\end{figure*}

The previous subsection fixed the centroid to isolate finite-source
interference. We now allow \(x_c(t)\) to vary, so the propagation distance
changes during emission. In Eq.~\eqref{eq:single}, this effect enters through
the factor \(e^{\mp ikx_c(t)}\), which combines with the temporal Fourier
phase to give \(e^{i\omega\theta_\pm(t)}\), where
\begin{equation}
\theta_{\pm}(t)\equiv t\mp\frac{x_c(t)}{c_L},
\qquad
\frac{d\theta_{\pm}}{dt}=1\mp\frac{\dot x_c}{c_L}
\label{eq:emission}
\end{equation}
For a centroid moving shoreward, successive emissions occur closer to a
shoreward observer and farther from a seaward observer, so \(\theta_+\) compresses the received history and \(\theta_-\) stretches it. The mapping remains one-to-one as long as \(1\mp\dot x_c/c_L>0\), which defines the
local regime considered below.

Over a shoreward leg with an approximately constant centroid velocity
\(\dot x_c=v>0\), a group-scale source oscillation at frequency \(f_g\)
is received at
\begin{equation}
f_{\pm}=\frac{f_g}{1\mp v/c_L}
\label{eq:doppler_periods}
\end{equation}
on the two branches. Shoreward arrivals are compressed
(\(f_+>f_g\)) and seaward arrivals are stretched (\(f_-<f_g\)): the
radiated infragravity field is Doppler-shifted by the migrating
breakpoint-forcing centroid (Figure~\ref{fig:doppler}).

A rigid footprint travelling with the incident group would have \(v=c_g\),
but the breakpoint-forcing centroid is a moment of a distributed dissipation
field. As the group grows and decays, breaking starts and stops at different
depths and redistributes the forcing within that field, so \(x_c\) need not
travel at \(c_g\) and can relocate offshore when the next group activates
breaking in deeper water.

Inserting \(x_c(t)=x_{c0}+vt\) and a fixed travel-time width
\(\Delta\tau_0\) into Eq.~\eqref{eq:single} transfers this arrival-time
rescaling to the source spectrum:
\begin{equation}
\big|\Rpm(\omega)\big|
=
\frac{1}{2c_L^2}
\big|\widehat\varphi(\omega\Delta\tau_0)\big|
\left|
\widehat A\!\left[
\omega\left(1\mp\frac{v}{c_L}\right)
\right]
\right|
\label{eq:doppler_spectrum}
\end{equation}
The first factor is the finite-width filter derived in
Section~\ref{subsec:lowpass}, and the second samples the temporal source
spectrum at a direction-dependent frequency. Where
\(\lvert\widehat A\rvert\) decreases with frequency, the shoreward branch
samples a lower source frequency for a given observed \(\omega\) and contains
more high-frequency energy, whereas the seaward branch samples a higher
frequency and loses more. Centroid migration can therefore counteract finite-width attenuation shoreward and reinforce it seaward.

For a nonuniform centroid trajectory, the arrival-time mapping
\(\theta_\pm(t)\) converts the changing centroid speed into a time-dependent
phase modulation, with Eq.~\eqref{eq:doppler_periods} as its local constant-velocity limit. Over a complete migration cycle, the two legs enter the spectrum with weights
set by \(A(t)\), so the leg with stronger breaking can dominate the long-record Doppler
signature even when the centroid returns to its initial position.

\subsection{Travel-time formulation over variable bathymetry}
\label{subsec:vardepth}

\begin{figure*}[pos=htbp]
\centering
\includegraphics[width=\textwidth]{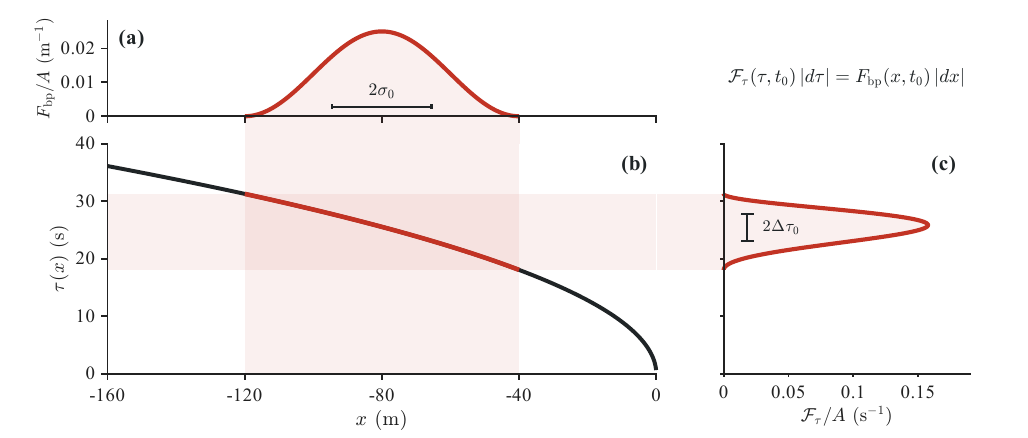}
\caption{One localised breakpoint-forcing footprint expressed in cross-shore
distance and long-wave travel time on a plane slope. (a) Spatial density.
(b) Change of coordinate $\tau(x)$. (c) Density per unit travel time.
Shaded bands follow the same part of the footprint through the coordinate
change; the reference descriptors map from
$(x_{c0},\sigma_0)$ to $(\tau_{c0},\Delta\tau_0)$.}
\label{fig:travel_time_mapping}
\end{figure*}

We now extend the analysis to variable bathymetry. Once $c_L$ varies with
depth, a single wavenumber can no longer describe propagation from every part
of the source to an observer. In the Green-function representation of
Eq.~\eqref{eq:green_rep}, each source position contributes with the
propagation phase and gain set by the bathymetry, including shoaling and
partial reflection. To expose the finite-width and migration effects within
that solution, we describe source position by long-wave travel time.

Let $x_{\mathrm{ref}}$ be a fixed reference position and define
\begin{equation}
\tau(x)
=
\int_x^{x_{\mathrm{ref}}}
\frac{dx'}{c_L(x')}
\label{eq:tau}
\end{equation}
Changing $x_{\mathrm{ref}}$ adds a constant to every source travel time,
shifting $\tau_c$ and the common phase origin without affecting
$\Delta\tau$, $\dot\tau_c$, or the physical response.

Since $d\tau/dx=-1/c_L$, the forcing density per unit travel time is
\begin{equation}
\begin{aligned}
&\mathcal{F}_\tau(\tau,t)
=
c_L\!\left[x(\tau)\right]
\Fbp\!\left[x(\tau),t\right]
\\
&\int\mathcal{F}_\tau(\tau,t)\,d\tau
=
A(t)
\end{aligned}
\label{eq:Ftau}
\end{equation}
Its centroid and root-mean-square width follow directly:
\begin{equation}
\begin{aligned}
&\tau_c(t)
=
\frac{1}{A(t)}
\int
\tau\,\mathcal{F}_\tau(\tau,t)\,d\tau
\\
&\Delta\tau^2(t)
=
\frac{1}{A(t)}
\int
\left[\tau-\tau_c(t)\right]^2
\mathcal{F}_\tau(\tau,t)\,d\tau
\end{aligned}
\label{eq:taumoments}
\end{equation}
The temporal weighting introduced in Eq.~\eqref{eq:dtau0} similarly defines
the reference centroids
\begin{equation}
\begin{aligned}
&x_{c0}
\equiv
\frac{\int A(t)x_c(t)\,dt}{\int A(t)\,dt}
\\
&\tau_{c0}
\equiv
\frac{\int A(t)\tau_c(t)\,dt}{\int A(t)\,dt}
\end{aligned}
\label{eq:centroid_refs}
\end{equation}

With the same moment normalisation used in cross-shore space, the source
becomes
\begin{equation}
\mathcal{F}_\tau(\tau,t)
\approx
\frac{A(t)}{\Delta\tau(t)}
\varphi_\tau\!\left(
\frac{\tau-\tau_c(t)}{\Delta\tau(t)}
\right)
\label{eq:Ftau_ansatz}
\end{equation}
where $\varphi_\tau$ has unit integral, zero mean, and unit variance. If
$c_L$ changes little across one forcing footprint, expanding $\tau(x)$ about
$x_c(t)$ relates the two source descriptions:
\begin{equation}
\begin{aligned}
&\tau_c(t)
\simeq
\tau\!\left[x_c(t)\right]
\\
&\Delta\tau(t)
\simeq
\frac{\sigma(t)}{c_L[x_c(t)]}
\\
&\varphi_\tau(\zeta)
\simeq
\varphi(-\zeta)
\end{aligned}
\label{eq:tau_local_map}
\end{equation}
The reversal follows because $\tau$ decreases shoreward and only conjugates
the transform of the reference shape, leaving
$\lvert\widehat\varphi\rvert$ unchanged.

Figure~\ref{fig:travel_time_mapping} distinguishes the travel time associated
with source position, which enters the Green response, from the spread of
travel times within the source, which controls interference through
$p=\omega\Delta\tau$. The local mapping in
Eq.~\eqref{eq:tau_local_map} therefore concerns only the footprint; the
exact Green function continues to describe propagation over the complete
bathymetry.

The connection with the constant-depth mechanisms follows by separating the
shoreward and seaward travelling components of the Green kernel, denoted by
$G_+$ and $G_-$, respectively. If the magnitude of $\partial_yG_\pm$
changes little across one footprint and its phase is locally represented by
$e^{\pm i\omega\tau(y)}$, inserting Eq.~\eqref{eq:Ftau_ansatz} into
Eq.~\eqref{eq:green_rep} gives
\begin{equation}
\begin{aligned}
\widehat\eta_\pm(x,\omega)
&\simeq
\int_{-\infty}^{\infty}
A(t)
\frac{\partial G_\pm}{\partial y}
\!\left(
x,x[\tau_c(t)];\omega
\right)
\\
&\qquad\times
\widehat\varphi_\tau\!\left[
\pm\omega\Delta\tau(t)
\right]
e^{i\omega t}\,dt
\end{aligned}
\label{eq:single_tau}
\end{equation}
where the upper and lower signs denote shoreward and seaward propagation,
respectively. The shape transform again measures cancellation across the
footprint through the phase spread $\omega\Delta\tau$, and the phase of the Green factor at the centroid combines with $e^{i\omega t}$ to form the
arrival-time coordinates $t\pm\tau_c(t)$. During shoreward migration,
$\dot\tau_c<0$, so successive emissions arrive closer together shoreward
and farther apart seaward. Thus $\Delta\tau$ and $-\dot\tau_c$ take the
roles played by $\sigma/c_L$ and $v/c_L$ on a horizontal bed. The full Green
function in Eq.~\eqref{eq:green_rep} also supplies the position-dependent
gain and reflected components.

Equation~\eqref{eq:single_tau} gives a local interpretation of the direct
travelling branches. The calculations below evaluate the complete Green
representation in Eq.~\eqref{eq:green_rep}, without freezing the kernel
across the forcing footprint. For the direct travelling path, a
Wentzel--Kramers--Brillouin (WKB) approximation gives the Green-function gain
for a compact source centred at $x_s$ and an observer at $x_o$ on a slowly
varying profile, with $c_{L,s}=c_L(x_s)$ and $c_{L,o}=c_L(x_o)$,
\begin{equation}
\begin{aligned}
&\left|
\frac{\partial G_\pm(x_o,x_s;\omega)}{\partial y}
\right|
\simeq
\frac{1}{2c_{L,o}^{1/2}c_{L,s}^{3/2}}
\\
&\left|\widehat\eta_\pm(x_o,\omega)\right|
\propto
\left|\widehat A(\omega)\right|
c_{L,o}^{-1/2}c_{L,s}^{-3/2}
\end{aligned}
\label{eq:wkb_depth_gain}
\end{equation}
The source-depth factor $c_{L,s}^{-3/2}$ controls the conversion of a fixed
integrated source into surface elevation, and $c_{L,o}^{-1/2}$ describes
the shoaling of the direct branch. In the travel-time formulation,
$\Delta\tau$ controls interference across the source and $-\dot\tau_c$
controls the direction-dependent compression and stretching of arrival
times. The calculations in Section~\ref{sec:xbeach_mechanisms} use the
complete Green function, including depth-dependent gain, partial reflection,
and departures from the WKB limit.

\section{Breakpoint-forced response across beach slopes}
\label{sec:xbeach_mechanisms}

\begin{figure*}[!t]
\centering
\includegraphics[width=\textwidth]{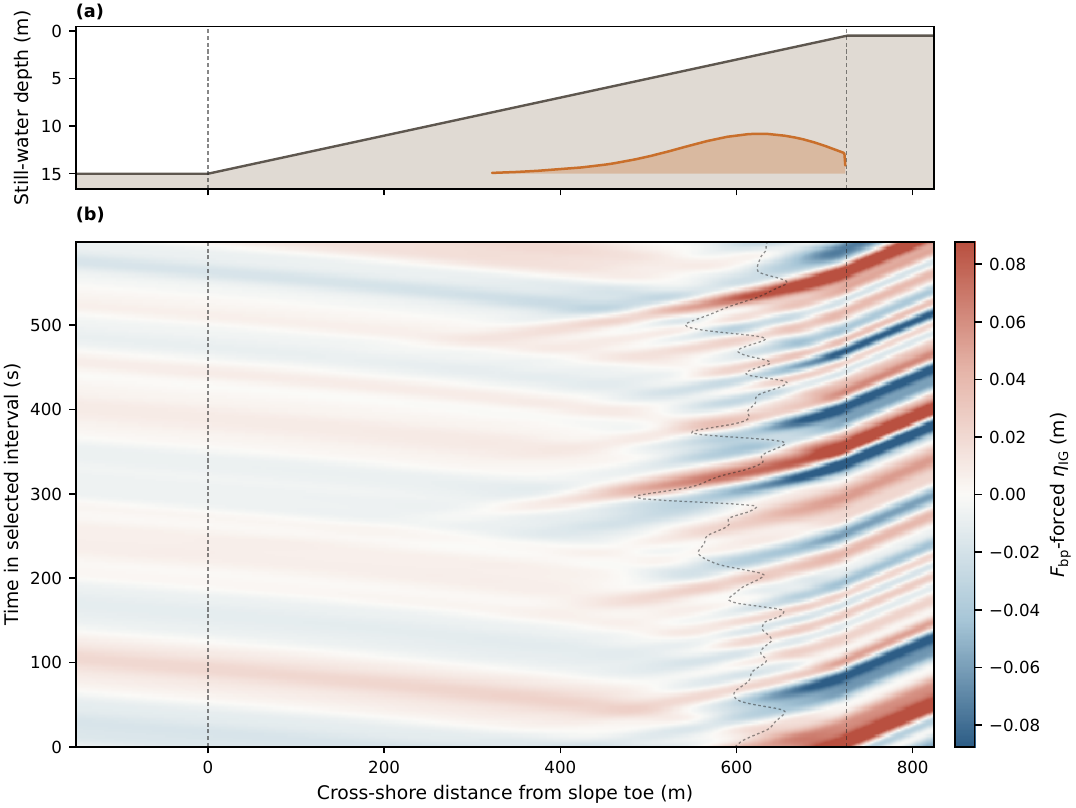}
\caption{Setup and breakpoint-forced radiation for the
\(1{:}50\) case. (a) Analysis bathymetry and mean group-scale primary
\(F_{\mathrm{bp}}\) profile, normalised and plotted within the bed.
(b) Green-function response to the complete diagnosed
\(F_{\mathrm{bp}}\) over \(600\)~s, band-limited to
\(0.004\leq f\leq0.05\)~Hz. The dotted curve marks the primary-source
centroid; the vertical dashed lines mark the slope toe and the
\(h=0.5\)~m source cutoff.}
\label{fig:section5_setup_radiation}
\end{figure*}

Besides the strength history \(A(t)\), Section~\ref{sec:green_filters}
identifies three source properties that shape the radiated waves: breaking
depth, source extent, and centroid migration. We
now quantify these controls across plane-beach slopes, using the
group-resolving model to diagnose \(F_{\mathrm{bp}}(x,t)\) and the
variable-depth Green solution to propagate it.

\subsection{Numerical experiment}
\label{subsec:section5_design}

Apart from the shoreline treatment, the simulations repeat the wave
conditions, breaking formulation, boundary forcing, and coupling choices of
Section~\ref{subsec:forcing_signatures}. Nine slopes from \(1{:}10\) to \(1{:}50\) in increments of five, covering
steep to intermediate natural beaches, follow a \(400\)-m-long, \(15\)-m-deep
approach and end in a shallow absorbing reach; each run uses
\(\Delta x=1\)~m and provides \(4\)~h of post-spin-up output at \(1\)-s
intervals.

After applying the \(0.05\)~Hz group-scale filter used in
Section~\ref{subsec:forcing_signatures}, we identify the connected footprint
around the dominant maximum. The complete filtered forcing drives the radiated responses; the primary
footprint provides the source descriptors and phase controls, and disconnected
patches form only a small residual.

On the plane slope, Eq.~\eqref{eq:green_def} has the Bessel basis
\(J_0(\chi)\) and \(Y_0(\chi)\), with
\(\chi=2\omega\sqrt{h/g}/|h_x|\) \citep{Liao2023}. Their Hankel combinations
are matched to outgoing waves in the adjoining constant-depth reaches. The
Green domain ends at \(h=0.5\)~m and the source is tapered near this cutoff,
retaining depth-dependent gain and partial reflection but excluding the nonlinear inner reach and shoreline returns.

\begin{figure}[pos=htbp]
\centering
\includegraphics[width=0.82\columnwidth]{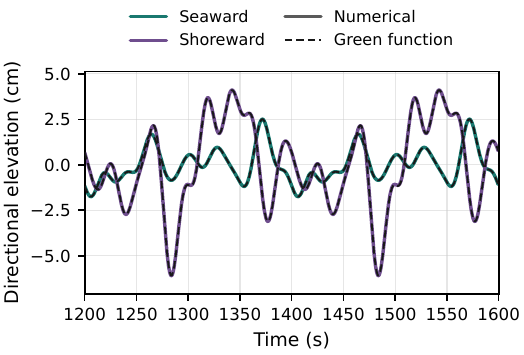}
\caption{Variable-depth Green validation on a \(1{:}20\) slope. Solid curves
show the linear long-wave model and dashed curves the Green solution for the
same multifrequency source.}
\label{fig:section5_green_validation}
\end{figure}

Figure~\ref{fig:section5_setup_radiation} shows the \(1{:}50\) geometry, mean
primary footprint, and \(600\)~s of radiation; panel~(b) resolves both
outgoing branches and the shoreward gain. A multifrequency Gaussian-source
check against the linear long-wave model gives overlapping solutions in
Figure~\ref{fig:section5_green_validation}, with amplitude differences below
one percent and phase differences below one degree on both branches.

All spectra use the complete \(4\)-h records and the same multitaper estimate;
spectral moments span \(0.004\leq f\leq0.05\)~Hz.

\subsection{Source properties and radiated response}
\label{subsec:section5_source_geometry}

Figure~\ref{fig:section5_source_geometry} connects beach slope to the local
breaking conditions and source geometry. From \(1{:}10\) to \(1{:}50\), the forcing-weighted source depth increases
from about \(2.8\) to \(3.5\)~m as the short-wave height decreases, and
\(\Delta\tau_0\) grows from about \(3\) to \(14\)~s. During active shoreward
migration, the \(A(t)\)-weighted median of \(v/c_L=-\dot\tau_c\) rises from
about \(0.13\) to \(0.30\), and its 90th percentile approaches \(0.5\) on the
gentler slopes. The value used in Figure~\ref{fig:doppler} therefore
represents a strong source leg within the diagnosed range.

To quantify the depth-dependent conversion of a prescribed breakpoint-forcing
source into surface elevation, we place the same compact source history from
the \(1{:}20\) case at depths between \(3\) and \(8\)~m, keeping \(A(t)\)
fixed. The shoreward response in panel~(d) follows the
\(h_s^{-3/4}\) direct-path scaling of Eq.~\eqref{eq:wkb_depth_gain}, whereas
partial reflection weakens the depth dependence of the seaward branch.

The complete diagnosed sources produce a clear contrast between the outgoing
branches in Figure~\ref{fig:section5_absolute_spectra}. Shoreward, the spectra
in panel~(b) keep nearly the same shape across slopes and change mainly in
level; seaward, panel~(a) shows an increasing concentration toward lower
frequencies as the beach becomes gentler. Accordingly, \(T_{m01,\mathrm{IG}}\) more than doubles seaward but changes
only modestly shoreward, and \(H_{m0,\mathrm{IG}}\) decreases on both branches. The following controls seek the origin of this directional contrast
in spectral shape.

\subsection{Separating finite-source interference and centroid migration}
\label{subsec:section5_controls}

The branch contrast in Figure~\ref{fig:section5_absolute_spectra} has the
directional structure expected from the two phase mechanisms derived in
Section~\ref{sec:green_filters}. Source extent creates a phase spread across
each forcing footprint, and centroid migration changes the spacing between
successive arrivals in opposite directions on the two branches. We test these
mechanisms directly in the diagnosed sources, first by quantifying the spatial
coherence of each footprint and then by constructing Green-function responses
in which one phase effect is removed at a time.

At frequency \(f\), contributions emitted at \(\tau\) and at the centroid
differ in phase by \(2\pi f[\tau-\tau_c(t)]\). Summing the diagnosed primary
field \(\mathcal{F}_{\mathrm p}(\tau,t)\), with source activity weighted by
\(A^2(t)\), gives the coherent power fraction
\begin{equation}
\mathcal{C}(f)
=
\frac{
\displaystyle\int A^2(t)
\left|
\frac{1}{A(t)}
\int
\mathcal{F}_{\mathrm p}(\tau,t)
e^{i2\pi f[\tau-\tau_c(t)]}
\,d\tau
\right|^2dt
}{
\displaystyle\int A^2(t)\,dt
}
\label{eq:section5_exact_coherence}
\end{equation}
The point-source limit gives \(\mathcal{C}=1\), so departures from unity
measure the loss of spatial coherence caused by finite source extent.

To determine how these phase effects enter the radiated spectra, let \(+\)
and \(-\) denote the shoreward and seaward branches and define
\(K_\pm(y,f)=\partial_yG_\pm(x_\pm,y;2\pi f)\). We take the diagnosed source
integral \(I_\pm\) as the reference response and construct two phase controls.
Spatial alignment shifts the propagation phase of every position within a
footprint to that of its centroid at the same instant, retaining the diagnosed
centroid trajectory. For the second control, centroid locking preserves the
internal phase structure of each footprint and replaces \(\tau_c(t)\) by its
onset value \(\tau_e\) during each active interval, with interpolation through
periods of weak source activity. The three source integrals are
\begin{equation}
\begin{aligned}
I_\pm(f,t)
&=
\int_{\mathcal S}
K_\pm(y,f)
F_{\mathrm{bp,p}}(y,t)\,dy
\\
I_\pm^{\mathrm{align}}(f,t)
&=
\int_{\mathcal S}
K_\pm(y,f)
F_{\mathrm{bp,p}}(y,t)
\\[-0.5ex]
&\quad{}\times
\exp\!\left(
\pm i\,2\pi f[\tau_c(t)-\tau(y)]
\right)\,dy
\\
I_\pm^{\mathrm{lock}}(f,t)
&=
I_\pm(f,t)
\exp\!\left[
\pm i\,2\pi f\{\tau_e-\tau_c(t)\}
\right]
\end{aligned}
\label{eq:section5_phase_lock}
\end{equation}
with corresponding outgoing amplitudes
\begin{equation}
\begin{aligned}
\widehat{\eta}_\pm(f)
&=
\int I_\pm(f,t)e^{i2\pi ft}\,dt
\\
\widehat{\eta}_\pm^{\mathrm{align}}(f)
&=
\int I_\pm^{\mathrm{align}}(f,t)e^{i2\pi ft}\,dt
\\
\widehat{\eta}_\pm^{\mathrm{lock}}(f)
&=
\int I_\pm^{\mathrm{lock}}(f,t)e^{i2\pi ft}\,dt
\end{aligned}
\label{eq:section5_controlled_amplitudes}
\end{equation}
Both controls apply unit-modulus phase corrections, so their comparison with
the diagnosed response preserves instantaneous source magnitude, source depth,
and Green-kernel gain. Spatial alignment becomes neutral in the point-source
limit, and centroid locking is neutral for a stationary centroid.

\begin{figure}[pos=p]
\centering
\includegraphics[width=\textwidth]{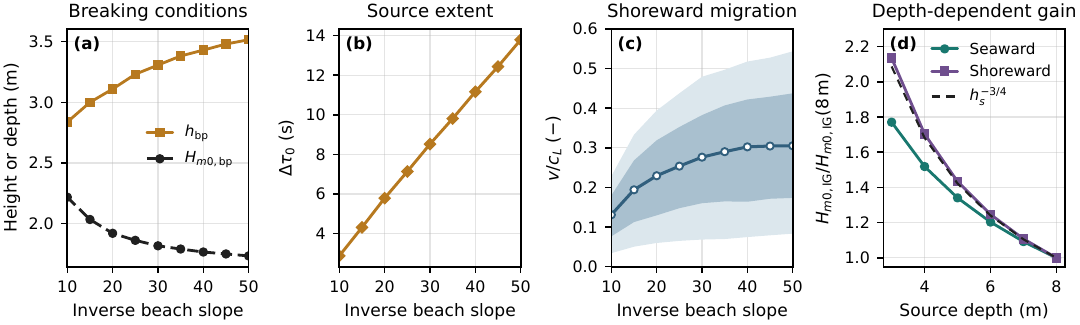}
\caption{Diagnosed source properties and depth-dependent response.
(a) Forcing-weighted source depth and corresponding short-wave height.
(b) Root-mean-square source width \(\Delta\tau_0\).
(c) \(A(t)\)-weighted median (line), interquartile range (dark shading), and
10th--90th percentile range (light shading) of \(v/c_L\) during active
shoreward migration. (d) Normalised response to the same compact source
history placed at different depths; the dashed curve gives the
\(h_s^{-3/4}\) direct-path scaling.}
\label{fig:section5_source_geometry}
\end{figure}

\begin{figure}[pos=p]
\centering
\includegraphics[width=0.94\textwidth]{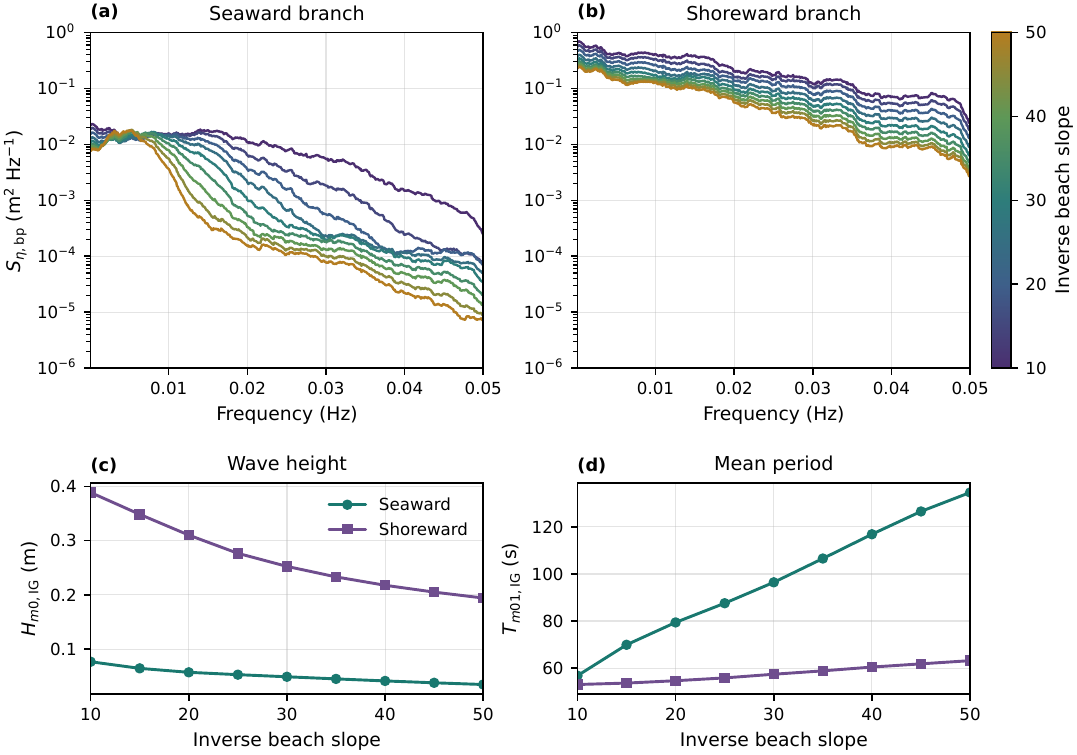}
\caption{Breakpoint-forced response from the complete diagnosed sources:
(a,b) seaward and shoreward spectra, (c) \(H_{m0,\mathrm{IG}}\), and
(d) \(T_{m01,\mathrm{IG}}\). Colours in (a,b) denote beach slope; curves in
(c,d) distinguish the two branches.}
\label{fig:section5_absolute_spectra}
\end{figure}

\begin{figure}[pos=p]
\centering
\includegraphics[width=\textwidth]{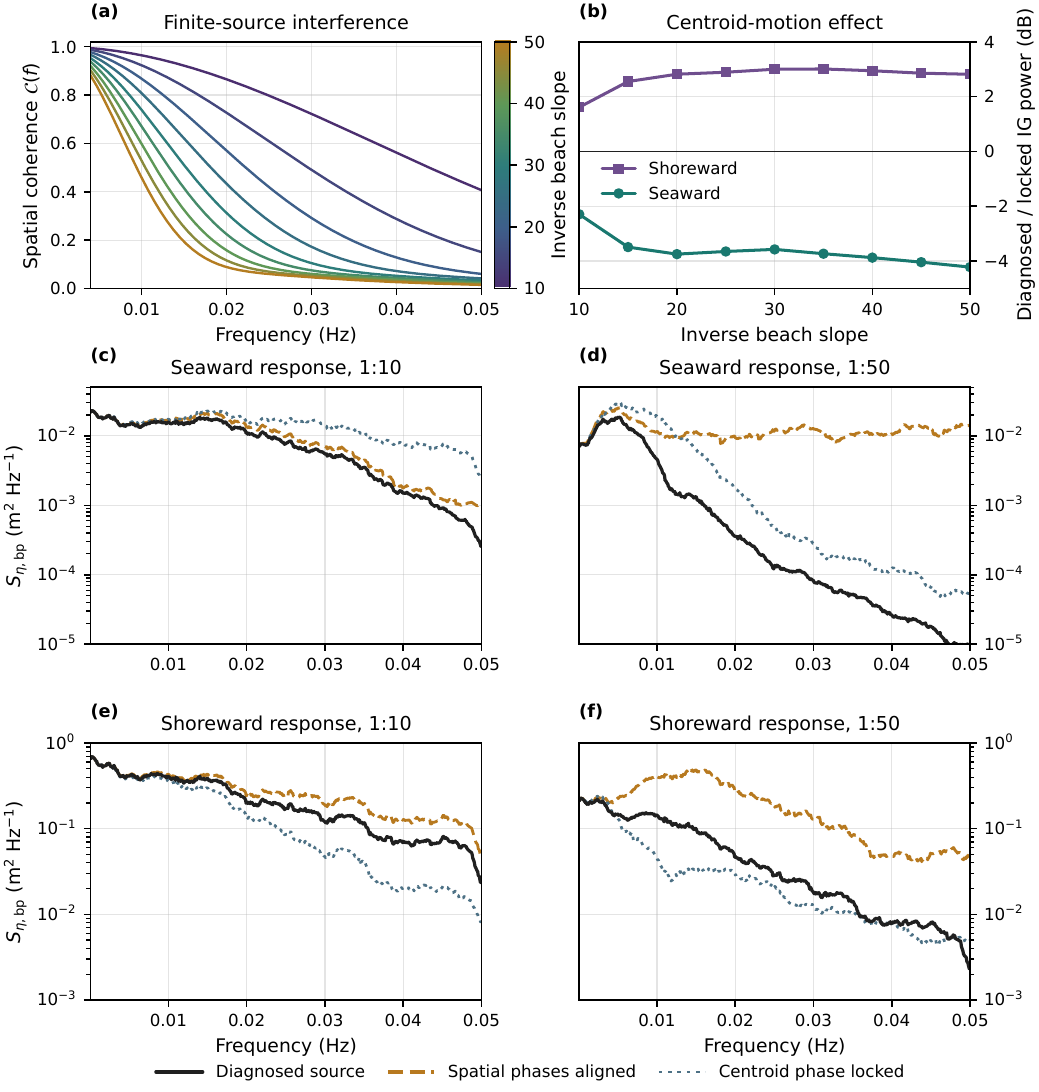}
\caption{Direct tests of finite-source interference and centroid migration.
(a) Source-power-weighted spatial coherence; colours denote beach slope.
(b) Ratio in dB of diagnosed to centroid-locked power over the infragravity
band. (c,d) Seaward and (e,f) shoreward spectra for \(1{:}10\) and
\(1{:}50\). Black curves use the diagnosed primary footprint, ochre curves
align its spatial phases, and blue curves lock its centroid phase within each
active event. Equations~\eqref{eq:section5_phase_lock}
and~\eqref{eq:section5_controlled_amplitudes} define the controls.}
\label{fig:section5_exact_mechanisms}
\end{figure}

Panel~(a) of Figure~\ref{fig:section5_exact_mechanisms} shows that spatial
coherence extends over a progressively narrower frequency range toward gentler
slopes, consistent with the increasing source extent in
Figure~\ref{fig:section5_source_geometry}. The phase-control experiment gives
the corresponding response-level test in panels~(c)--(f): aligning the spatial
phases restores high-frequency power on both branches, with a much stronger
effect for the broader \(1{:}50\) source. These results
confirm that the increasing upper-band attenuation toward gentler slopes is
produced by destructive interference across the forcing footprint.

Centroid locking tests the second analytical prediction. The diagnosed
response lies above the locked response shoreward and below it seaward,
matching the arrival-time compression and stretching derived in
Section~\ref{subsec:doppler}; panel~(b) shows that the same directional effect persists across all nine slopes. It also survives averaging over the
full records because stronger source activity is preferentially associated
with shoreward migration, whereas much of the offshore motion occurs during
weaker forcing.

The two controls therefore reproduce the analytical signatures of finite-source
interference and centroid migration under irregular waves and variable
bathymetry. As the source broadens toward gentler slopes, finite-source
interference increasingly suppresses upper-band energy on both branches;
centroid migration strengthens this spectral shift seaward and compensates
part of it shoreward. Their combination accounts for the strong seaward
spectral lengthening and the comparatively stable shoreward spectral shape in
Figure~\ref{fig:section5_absolute_spectra}.

\onecolumn

\clearpage

\clearpage
\twocolumn

\section{Discussion}
\label{sec:discussion}

The results of Sections~\ref{sec:rs-split-short}
to~\ref{sec:xbeach_mechanisms} bear on the following questions: how the
present separation relates to the vortex-force proxy of
\citet{Midouni2025CD} (Section~\ref{subsec:link_vortex_force}) and to the
moving-breakpoint model of \citet{Symonds1982}
(Section~\ref{subsec:link_symonds1982}), why migration of the breaking
region affects the two branches differently
(Section~\ref{subsec:doppler_directionality}), how the breakpoint-forced
response depends on beach slope
(Section~\ref{subsec:discussion_controls_amplitude}), what this implies
for reefs and other steep profiles (Section~\ref{subsec:reefs}), and how
far the results extend (Section~\ref{subsec:limitations}).

\subsection{Relation to the vortex-force separation}
\label{subsec:link_vortex_force}

The present decomposition builds on the separation proposed by
\citet{Midouni2025CD} within the vortex-force formalism. That formulation
evolves the quasi-Eulerian velocity, so wave effects appear in several
places: the Bernoulli head and the breaking term in the momentum equation,
and the divergence of the wave-induced mass transport in the continuity
equation. Breakpoint forcing was identified with the breaking term alone,
and the other two were removed to isolate its response.

This term-by-term choice is ambiguous, because the effect of dissipation on
the mean flow is not confined to the term in which \(D_b\) appears
explicitly. The breaking term represents the momentum released directly by
the waves, \(D_b/(\rho c)\) in transport form, whereas the total
wave-induced forcing responds to dissipation through the radiation stress,
by \(\alpha D_b/(\rho c_g)\), which coincides with the former in deep water
but exceeds it by half in shallow water.

The total-transport formulation used here removes the ambiguity. All wave
effects enter the momentum balance through a single wave-induced flux
gradient, so that a flow without this gradient carries no wave-induced
forcing at all, and the short-wave energy balance partitions the gradient
exactly into a part proportional to \(D_b\), which
Eq.~\eqref{eq:full_Fbp} identifies as \(F_{\mathrm{bp}}\), and the rest,
\(F_{\mathrm{bw}}\), which collects the local change of wave action and the
variations of the wave coefficients along the path.

\subsection{Relation to the moving-breakpoint model}
\label{subsec:link_symonds1982}

The present framework also generalises the forcing parametrization of
\citet{Symonds1982}, who first isolated long-wave generation by a
time-varying breakpoint. Their saturation relation prescribes the short
waves shoreward of the breakpoint and confines the time-varying forcing to
the region swept by its motion, so that the extent of the forcing region
and its motion are both set by the prescribed breakpoint trajectory.

That switched forcing is, moreover, not pure breakpoint forcing in the
present sense. For saturated waves the radiation stress is steady, so
\(F_{\mathrm{bw}}\) reduces to the depth-gradient term of
Eq.~\eqref{eq:forcing_split_defs}, and in shallow water the saturated
radiation-stress gradient splits into \(F_{\mathrm{bp}}\) and
\(F_{\mathrm{bw}}\) equal to \(5/4\) and \(-1/4\) of its value,
respectively. The forcing of \citet{Symonds1982} thus combines breakpoint
forcing with a smaller, opposing bound-wave contribution, and omits the
modulated forcing seaward of the breakpoint altogether.

Here, the starting point is instead a general spatiotemporal dissipation
\(D_b(x,t)\), from which the associated breakpoint forcing is diagnosed
without prescribing a breakpoint trajectory or a saturated surf zone. Its
distributed nature then becomes part of the problem: the location--scale
representation separates changes in strength, position, and extent, and
the Green-function analysis shows how finite-source interference and
centroid migration shape the radiated waves separately.

\subsection{Directionality of the Doppler effect}
\label{subsec:doppler_directionality}

The directional Doppler effect calls for a closer look, because the theory
of Section~\ref{sec:green_filters} gives the shift for one leg of the
migration whereas the problem is periodic by nature. The breaking region
moves back and forth with the groups, so the compression produced on one
leg would be undone by the stretching on the other, leaving no net
difference between the branches, if the forcing were the same on both legs.

The asymmetry comes from the largest waves of a group, which break farthest
offshore and keep dissipating on their way to the coast. The strongest
forcing therefore accompanies a breaking region moving shoreward, whereas
the jump back offshore happens when the next group starts breaking, with
little dissipation yet. The net effect over a record is set by this
systematic pairing of strong forcing with shoreward motion, repeated from
group to group, and is expected to favour the shoreward branch.
Section~\ref{subsec:section5_controls} confirms both steps: the diagnosed
histories pair the strongest forcing with shoreward migration, and the
centroid-locking control shows the resulting gain of the shoreward spectrum
and loss of the seaward one.

\subsection{Slope dependence of the breakpoint-forced response}
\label{subsec:discussion_controls_amplitude}

Previous assessments of breakpoint forcing have been mostly comparative:
its importance is judged against bound-wave release, through the
normalised bed slope \citep{Battjes2004} and the surf beat similarity
parameter \citep{Baldock2012}, most recently by \citet{Contardo2025}, who
isolated each mechanism in a linear model with a fixed breakpoint, found
the efficiency of breakpoint forcing independent of slope, and attributed
the transition in dominance to bound-wave release. Such a comparison is
demanding, because the two contributions are of a different nature:
\(F_{\mathrm{bp}}\) acts where and when short waves dissipate, whereas the
response to \(F_{\mathrm{bw}}\) integrates the whole shoaling history, with
the bound wave, the free waves generated over varying depth, and their
release adding to the breakpoint-forced waves with phase differences that
reduce the total \citep{Schaeffer1993}.

The separation of Section~\ref{sec:rs-split-short} allows breakpoint
forcing to be examined on its own for any dissipation \(D_b(x,t)\), and
hence with the finite extent and the migration that breaking has in
irregular waves. Under these conditions the breakpoint-forced response has
a slope dependence of its own: in the simulations of
Section~\ref{sec:xbeach_mechanisms}, a gentler slope gives a deeper, wider,
and more mobile source, and a weaker radiation concentrated at lower
frequencies. This result agrees with the earlier comparisons and, if
anything, strengthens them: on mild slopes, breakpoint forcing loses ground
not only because bound-wave release becomes more efficient, but also
because its own radiation weakens.

\subsection{Implications for reefs and steep profiles}
\label{subsec:reefs}

Coral reefs are the setting in which breakpoint forcing has most often been
reported to dominate \citep{Pomeroy2012, Buckley2018, Masselink2019}. Our
results point to a reason intrinsic to breakpoint forcing. Over a steep
fore reef, waves break in relatively shallow water and within a narrow
region, so that the forcing is efficiently converted into surface
elevation and suffers little interference across the infragravity band.
The response then depends mainly on the strength of the forcing and on the
breaking depth, its spatial structure entering only through the small
correction of Eq.~\eqref{eq:coherent_expansion}. On a steep profile the two
go together, since breaking occurs in shallower water and at greater wave
height.

Two field studies support this picture. \citet{Becker2016} reproduced most
of the low-frequency variance at the shoreline of steep Pacific fringing
reefs using the dynamics of \citet{Symonds1982} with breakpoint forcing
concentrated at the reef edge and linear damping over the flat, which is
consistent with this insensitivity to the structure of the forcing. The
observations of \citet{Liu2023}, with breakpoint forcing dominant although
the waves broke over a horizontal reef flat, fit the same interpretation:
after a steep fore reef the waves break in shallow water over a limited
region, whatever the local slope at the breakpoint. Their reading of this
result, as a dependence on the preceding cross-shore evolution of the short
waves, is what the source properties make explicit.

\subsection{Scope and limitations}
\label{subsec:limitations}

The results of Section~\ref{sec:xbeach_mechanisms} were obtained with a
single breaking closure and a single offshore sea state. The mechanisms of
Section~\ref{sec:green_filters} are independent of this choice, since they
hold for any prescribed \(F_{\mathrm{bp}}\) and the phase controls tested
them directly on the diagnosed sources. The slope trends of the source
should hold beyond this particular setting as well, at least
qualitatively, since they follow from the geometry of the profile: on a
gentler slope the same depth range spans a greater distance, so breaking
spreads over a wider region and the breakpoint excursion grows, and the
waves lose more energy before reaching a given depth, which shifts the
dissipation toward deeper water.

The magnitudes reached at a given slope, and the degree to which the
forcing concentrates on the shoreward leg, are more sensitive, because they
depend on how the parametrization distributes dissipation within a group.
A surface roller, in particular, would delay and spread the effective
forcing, since roller energy travels shoreward before transferring its
momentum to the flow. The influence of wave height, period, and spectral
width remains to be explored. More fundamentally, the mechanisms themselves
still await confirmation outside the group-resolving framework, by
phase-resolving simulations, laboratory experiments, or in situ
observations.

The complete nearshore response also contains the shoaling bound wave
driven by \(F_{\mathrm{bw}}\) and its release at breaking, whose relative
importance increases toward milder slopes. The released and
breakpoint-forced components are in antiphase on each branch in constant
depth (Eq.~\eqref{eq:ratio_green}), and their coherent sum may reshape the
trends found here for breakpoint forcing alone. The corresponding
Green-function solution for the bound-wave forcing is available
\citep{Liao2023}, and combining the two solutions would allow both
contributions to be followed together over variable bathymetry.

The Green-function analysis of Sections~\ref{sec:green_filters}
and~\ref{sec:xbeach_mechanisms} treats the long-wave dynamics as linear, an
approximation that degrades toward the shore, where infragravity
elevations become a sizeable fraction of the depth. The resulting nonlinear
effects \citep{Rijnsdorp2022} alter the signal recorded near the shore but
are expected to leave the main conclusions of this study unchanged.

Finally, the calculations assume one-dimensional, normally incident waves,
and leave aside alongshore variability, the effect of currents on the short
waves, and bottom friction.

\section{Conclusion}

In this study, we sought to better understand the generation of infragravity
waves by breakpoint forcing. The first step was to find a way to isolate this
mechanism in the governing equations. The short-wave energy balance provides
one: the radiation stress gradient splits into a part proportional to the
local breaking dissipation, which we identify as breakpoint forcing, and a
bound-wave forcing part. In constant depth, the two contributions behave as
expected, with breakpoint forcing radiating equally in both directions, the
released bound wave mainly shoreward, and the two in antiphase. We then
implemented the separation in a one-dimensional group-resolving model and
compared it with SWASH on a plane beach. The two models agree on the main
features of the long-wave field, and the separation shows which mechanism
generates each of them.

We then focused exclusively on breakpoint forcing, starting with the
structure of \(D_b\), and hence of \(F_{\mathrm{bp}}\), at the group scale.
The forcing is confined to the breaking region and evolves with the passage
of each group: breaking starts farther offshore as the largest waves of the
group arrive, is strongest while the breaking region shifts shoreward, then
weakens and contracts until the next group starts breaking farther offshore
again. Its strength, position and width therefore all vary within a group
cycle, which led us to represent the forcing as a location--scale family: a
fixed shape, scaled, translated and stretched over time by these three
quantities. This representation was found to reproduce closely the forcing
diagnosed in the group-resolving model.

We then solved the long-wave equation with a Green function approach,
inserting this location--scale representation of the forcing, which yielded
three main results. The first is a finite-source filtering: because the
breaking region has a finite cross-shore width, the waves emitted from its
different parts cancel out at the shortest infragravity periods, so that
generation itself acts as a low-pass filter that reaches lower frequencies as
the region widens. The second is a directional Doppler effect: as the
breaking region migrates, the waves radiated shoreward are compressed in time
and those radiated seaward are stretched, which, since breaking is strongest
during shoreward migration, partly compensates the filtering shoreward and
reinforces it seaward. The third is the effect of the breaking depth, which
sets how efficiently the forcing is converted into surface elevation,
shallower breaking giving a stronger response.

These predictions were then tested with numerical experiments on plane
beaches of various slopes, in which the forcing diagnosed in the
group-resolving model was propagated with the Green function solution. As the
slope decreases, waves break in deeper water over a wider and more mobile
region, and the breakpoint-forced infragravity waves become weaker and
longer, mostly in the seaward direction. Three controlled tests confirmed
where these changes come from: removing the phase differences across the
breaking region restored the high frequencies in both directions, freezing
its migration suppressed the asymmetry between the shoreward and seaward
waves, and moving the same forcing to different depths gave the predicted
depth dependence.

These results give coastal scientists and modellers a definition of
breakpoint forcing that group-resolving models can implement directly,
together with a new picture of the mechanism itself, as the radiation of a
breaking region of finite width that moves with the wave groups. This picture
predicts the shape of the infragravity spectra radiated on either side of the
breaking region, and thereby gives a physical basis for better interpreting
nearshore measurements and simulations.

\printcredits

\section*{Acknowledgements}
This research was funded by the FWO Infragravity Waves project (3E221057) and was partly supported by the Bijzonder Onderzoeksfonds (BOF) of KU Leuven under the Startfinanciering programme (Project ID: 3E241133, Reference: STG/24/026). The authors thank Robert McCall (Deltares) for help with model development and discussions during the early stages of this work, and Xavier Bertin (LIENSs, CNRS) for earlier discussions that were instrumental in its genesis.

\section*{Data availability}
Data will be made available upon request.

\appendix
\numberwithin{equation}{section}
\renewcommand{\thefigure}{\thesection\arabic{figure}}
\renewcommand{\theHfigure}{\thesection\arabic{figure}}
\setcounter{figure}{0}

\newpage

\section{Coupled one-dimensional wave--flow system and forcing separation}
\label{app:full_1d_system}

We extend the reduced system of Section~\ref{sec:rs-split-short} by including
the instantaneous free surface and the quasi-Eulerian current in the
short-wave equations. Action conservation \citep{BrethertonGarrett1968} and
crest conservation \citep{Uchiyama2010,Marchesiello2026} are coupled to the
total-transport equations of \citet{Smith2006}.

\subsection{Governing equations}

Let $x$ increase shoreward and $d=h+\eta$ be the instantaneous depth. The
action density $\mathcal{N}$ and the wavenumber $k_w$ satisfy
\begin{subequations}
\begin{align}
&\frac{\partial\mathcal{N}}{\partial t}
+\frac{\partial}{\partial x}
\left(V_g\mathcal{N}\right)
=-\frac{D_b}{\sigma_w}
\label{eq:full_action}
\\
&\frac{\partial k_w}{\partial t}
+\frac{\partial}{\partial x}
\left(U_E k_w+\sigma_w\right)
=0
\label{eq:full_crest_conservation}
\end{align}
\end{subequations}
where $D_b$ is the depth-induced breaking dissipation, $V_g=U_E+c_g$ the
absolute group velocity, and $\sigma_w$, $c$ and $c_g$ follow from
\begin{equation}
\sigma_w^2=gk_w\tanh(k_w d),
\qquad
c=\frac{\sigma_w}{k_w},
\qquad
c_g=\frac{\partial\sigma_w}{\partial k_w}
\label{eq:full_dispersion}
\end{equation}
The wave energy is $E=\sigma_w\mathcal{N}$, the wave momentum
$M^W=\mathcal{N}k_w/\rho=E/(\rho c)$, and $S_{xx}=\alpha E$ with $\alpha$
given by Eq.~\eqref{eq:Sxx-def}. With $q$ the total transport, continuity
and momentum read
\begin{subequations}
\begin{align}
&\frac{\partial d}{\partial t}
+\frac{\partial q}{\partial x}
=0
\label{eq:full_continuity} \\
&\frac{\partial q}{\partial t}
+\frac{\partial}{\partial x}
\left[
\frac{q^2}{d}
+\frac{gd^2}{2}
+\frac{S_{xx}}{\rho}
-\frac{(M^W)^2}{d}
\right]
=gd\frac{\partial h}{\partial x}
\label{eq:full_momentum}
\end{align}
\end{subequations}
The wave term in Eq.~\eqref{eq:full_momentum} is the excess momentum flux of
\citet{Phillips1977}, $S_{xx}-\rho(M^W)^2/d$, because the advective flux
$q^2/d$ of the total transport already contains the flux carried by the wave
momentum \citep{Smith2006}. The quasi-Eulerian current entering the
short-wave equations is $U_E=(q-M^W)/d$.

\subsection{Forcing separation}

The wave terms in Eq.~\eqref{eq:full_momentum} define the total wave-induced
forcing
\begin{equation}
F_{\mathrm{wave}}
=-\frac{1}{\rho}
\frac{\partial}{\partial x}
\left[
S_{xx}-\rho\frac{(M^W)^2}{d}
\right]
\label{eq:full_wave_force}
\end{equation}
Solving Eq.~\eqref{eq:full_action} for the spatial action gradient,
\begin{equation}
\frac{\partial\mathcal{N}}{\partial x}
=-\frac{1}{V_g}
\left(
\frac{D_b}{\sigma_w}
+\frac{\partial\mathcal{N}}{\partial t}
+\mathcal{N}\frac{\partial V_g}{\partial x}
\right)
\label{eq:full_action_gradient}
\end{equation}
and substituting into Eq.~\eqref{eq:full_wave_force} with
$E=\sigma_w\mathcal{N}$ and $M^W=\mathcal{N}k_w/\rho$ separates the forcing
into a part proportional to the breaking dissipation and the rest:
\begin{equation}
\begin{aligned}
F_{\mathrm{bp}}
&=\frac{\beta}{\rho V_g}\,D_b,
\qquad
\beta\equiv\alpha-\frac{2M^W}{cd}
\\
F_{\mathrm{bw}}
&=\frac{\beta\sigma_w}{\rho V_g}\frac{\partial\mathcal{N}}{\partial t}
-\frac{S_{xx}}{\rho}\frac{\partial}{\partial x}
\ln\frac{\alpha\sigma_w}{V_g}
+\frac{(M^W)^2}{d}\frac{\partial}{\partial x}
\ln\frac{k_w^2}{V_g^2\,d}
\end{aligned}
\label{eq:full_Fbp}
\end{equation}
We identify $F_{\mathrm{bp}}$ with breakpoint forcing. The three terms of
$F_{\mathrm{bw}}$ are the local change of wave action and the variations of
the wave coefficients with depth and current along the path, for the
radiation stress and for the wave-momentum flux respectively; their long-wave
response is the bound wave and the free waves generated as it shoals and is
released. The terms in $M^W$ are of fourth order in the wave amplitude, and
dropping them with $V_g=c_g$, constant $\sigma_w$ and still-water
coefficients gives back Eq.~\eqref{eq:forcing_split_defs}.

The decomposition holds at each point and instant but requires $V_g\neq0$,
since the action gradient is divided by the absolute group velocity: the
total forcing remains well defined everywhere, whereas the two components
grow without bound where $V_g$ vanishes. This happens close to the
shoreline, where the long-wave velocity becomes comparable to $c_g$ and
$V_g$ can pass through zero during the seaward phase of the infragravity
motion. This is the blocking point of linear wave theory, at which the waves
actually break or are reflected \citep{ShyuPhillips1990, ChawlaKirby2002}, a
regime that wave-averaged models do not represent; XBeach, for instance,
applies wave--current interaction only above a minimum depth
\citep{Roelvink2009}. We likewise leave the current out of the short-wave
equations: in Sections~\ref{subsec:forcing_signatures}
and~\ref{sec:xbeach_mechanisms}, the short waves feel the instantaneous
depth $d=h+\eta$, but $U_E$ is set to zero in
Eqs.~\eqref{eq:full_action} and~\eqref{eq:full_crest_conservation}, in the
breaking closure and in $F_{\mathrm{bp}}$, so that $V_g=c_g>0$ everywhere;
since $U_E$ remains small compared with $c_g$ in the breaking region, this
has a limited effect on the forcing there.

Finally, the responses to $F_{\mathrm{bp}}$ and $F_{\mathrm{bw}}$ add up to
the response to the total forcing only for prescribed short waves and linear
long-wave dynamics; in the coupled nonlinear runs of
Section~\ref{subsec:forcing_signatures}, a calculation forced by one term
alone develops its own wave and flow state, so the component responses are
diagnostics and need not sum exactly to the complete solution.

\subsection{Numerical implementation}

The nonlinear shallow-water equations use a staggered C-grid, with $\eta$ and
$d=h+\eta$ at cell centres and $q$ at faces. We follow the conservative
mass-flux formulation of \citet{StellingDuinmeijer2003}, with the second-order
MinMod reconstruction of \citet{MihamiRoeber2026}.
Crest conservation uses a local Lax--Friedrichs Hamiltonian with Koren
one-sided phase gradients \citep{Koren1993}; the same reconstruction supplies
wave-action face states. A two-stage strong-stability-preserving Runge--Kutta
method advances both short-wave equations. Strang splitting integrates
breaking around transport, and a second-order partitioned Runge--Kutta scheme
couples short and long waves.
\section*{Declaration of generative AI and AI-assisted technologies in the
manuscript preparation process}
During the preparation of this work, the authors used Claude (Anthropic),
ChatGPT and Codex (OpenAI) to assist with the development and debugging of
the numerical model and analysis code, with literature searches, and with the
language and readability of the text. The authors reviewed and edited the
output as needed and take full responsibility for the content of the
published article.\bibliographystyle{cas-model2-names}
\bibliography{cas-refs}

@article{Baldock2000,
  author  = {Baldock, T. E. and Huntley, D. A. and Bird, P. A. D. and O'Hare, T. and Bullock, G. N.},
  title   = {Breakpoint generated surf beat induced by bichromatic wave groups},
  journal = {Coastal Engineering},
  year    = {2000},
  volume  = {39},
  number  = {2--4},
  pages   = {213--242},
  doi     = {10.1016/S0378-3839(99)00061-7}
}

@article{Baldock2006,
  author  = {Baldock, T. E.},
  title   = {Long wave generation by the shoaling and breaking of transient wave groups on a beach},
  journal = {Proceedings of the Royal Society A: Mathematical, Physical and Engineering Sciences},
  year    = {2006},
  volume  = {462},
  number  = {2070},
  pages   = {1853--1876},
  doi     = {10.1098/rspa.2005.1642}
}

@article{Baldock2010,
  author  = {Baldock, T. E. and Manoonvoravong, P. and Pham, K. S.},
  title   = {Sediment transport and beach morphodynamics induced by free long waves, bound long waves and wave groups},
  journal = {Coastal Engineering},
  year    = {2010},
  volume  = {57},
  number  = {10},
  pages   = {898--916},
  doi     = {10.1016/j.coastaleng.2010.05.006}
}

@article{Baldock2012,
  author  = {Baldock, T. E.},
  title   = {Dissipation of incident forced long waves in the surf zone - Implications for the concept of ``bound'' wave release at short wave breaking},
  journal = {Coastal Engineering},
  year    = {2012},
  volume  = {60},
  pages   = {276--285},
  doi     = {10.1016/j.coastaleng.2011.11.002}
}

@article{BaldockHuntley2002,
  author  = {Baldock, T. E. and Huntley, D. A.},
  title   = {Long-wave forcing by the breaking of random gravity waves on a beach},
  journal = {Proceedings of the Royal Society A: Mathematical, Physical and Engineering Sciences},
  year    = {2002},
  volume  = {458},
  number  = {2025},
  pages   = {2177--2201},
  doi     = {10.1098/rspa.2002.0962}
}

@article{Battjes2004,
  author  = {Battjes, J. A. and Bakkenes, H. J. and Janssen, T. T. and {van Dongeren}, A. R.},
  title   = {Shoaling of subharmonic gravity waves},
  journal = {Journal of Geophysical Research: Oceans},
  year    = {2004},
  volume  = {109},
  number  = {C2},
  pages   = {C02009},
  doi     = {10.1029/2003JC001863}
}

@article{Becker2016,
  title = {Infragravity waves on fringing reefs in the tropical Pacific: Dynamic setup},
  volume = {121},
  ISSN = {2169-9291},
  url = {http://dx.doi.org/10.1002/2015JC011516},
  DOI = {10.1002/2015jc011516},
  number = {5},
  journal = {Journal of Geophysical Research: Oceans},
  publisher = {American Geophysical Union (AGU)},
  author = {Becker,  J. M. and Merrifield,  M. A. and Yoon,  H.},
  year = {2016},
  month = May,
  pages = {3010–3028}
}

@article{Bertin2016,
  author  = {Bertin, Xavier and Olabarrieta, Maitane},
  title   = {Relevance of infragravity waves in a wave-dominated inlet},
  journal = {Journal of Geophysical Research: Oceans},
  year    = {2016},
  volume  = {121},
  number  = {8},
  pages   = {5418--5435},
  doi     = {10.1002/2015JC011444}
}

@article{Bertin2018,
  author  = {Bertin, Xavier and {de Bakker}, Anouk T. M. and {van Dongeren}, Ap and Coco, Giovanni and Andr{\'e}, Gael and Ardhuin, Fabrice and Bonneton, Philippe and Bouchette, Fr{\'e}d{\'e}ric and Castelle, Bruno and Crawford, Wayne C. and Davidson, Mark and Deen, Martha and Dodet, Guillaume and Gu{\'e}rin, Thomas and Inch, Kris and Leckler, Fabien and McCall, Robert and Muller, H{\'e}lo{\"\i}se and Olabarrieta, Maitane and Roelvink, Dano and Ruessink, Gerben and Sous, Damien and Stutzmann, {\'E}l{\'e}onore and Tissier, Marion F. S.},
  title   = {Infragravity waves: From driving mechanisms to impacts},
  journal = {Earth-Science Reviews},
  year    = {2018},
  volume  = {177},
  pages   = {774--799},
  doi     = {10.1016/j.earscirev.2018.01.002}
}

@article{Bertin2020,
  author  = {Bertin, Xavier and Martins, K{\'e}vin and {de Bakker}, Anouk and Chataigner, Teddy and Gu{\'e}rin, Thomas and Coulombier, Thibault and {de Viron}, Olivier},
  title   = {Energy Transfers and Reflection of Infragravity Waves at a Dissipative Beach Under Storm Waves},
  journal = {Journal of Geophysical Research: Oceans},
  year    = {2020},
  volume  = {125},
  number  = {5},
  pages   = {e2019JC015714},
  doi     = {10.1029/2019JC015714}
}

@article{Bowers1977,
  author  = {Bowers, E. C.},
  title   = {Harbour resonance due to set-down beneath wave groups},
  journal = {Journal of Fluid Mechanics},
  year    = {1977},
  volume  = {79},
  number  = {1},
  pages   = {71--92},
  doi     = {10.1017/S0022112077000044}
}

@article{BrethertonGarrett1968,
  author  = {Bretherton, F. P. and Garrett, C. J. R.},
  title   = {Wavetrains in inhomogeneous moving media},
  journal = {Proceedings of the Royal Society of London. Series A, Mathematical and Physical Sciences},
  year    = {1968},
  volume  = {302},
  number  = {1471},
  pages   = {529--554},
  doi     = {10.1098/rspa.1968.0034}
}

@article{Buckley2018,
  title = {Mechanisms of Wave‐Driven Water Level Variability on Reef‐Fringed Coastlines},
  volume = {123},
  ISSN = {2169-9291},
  url = {http://dx.doi.org/10.1029/2018JC013933},
  DOI = {10.1029/2018jc013933},
  number = {5},
  journal = {Journal of Geophysical Research: Oceans},
  publisher = {American Geophysical Union (AGU)},
  author = {Buckley,  M. L. and Lowe,  R. J. and Hansen,  J. E. and van Dongeren,  A. R. and Storlazzi,  C. D.},
  year = {2018},
  month = May,
  pages = {3811–3831}
}

@book{Casella2002,
  author    = {Casella, George and Berger, Roger L.},
  title     = {Statistical Inference},
  edition   = {2},
  publisher = {Duxbury},
  year      = {2002},
  isbn      = {0534243126}
}

@article{ChawlaKirby2002,
  author  = {Chawla, Arun and Kirby, James T.},
  title   = {Monochromatic and random wave breaking at blocking points},
  journal = {Journal of Geophysical Research: Oceans},
  year    = {2002},
  volume  = {107},
  number  = {C7},
  pages   = {3067},
  doi     = {10.1029/2001JC001042}
}

@article{Cheriton2016,
  author  = {Cheriton, Olivia M. and Storlazzi, Curt D. and Rosenberger, Kurt J.},
  title   = {Observations of wave transformation over a fringing coral reef and the importance of low-frequency waves and offshore water levels to runup, overwash, and coastal flooding},
  journal = {Journal of Geophysical Research: Oceans},
  year    = {2016},
  volume  = {121},
  number  = {5},
  pages   = {3121--3140},
  doi     = {10.1002/2015JC011231}
}

@article{Contardo2021,
  author  = {Contardo, Stephanie and Lowe, Ryan J. and Hansen, Jeff E. and Rijnsdorp, Dirk P. and Dufois, Fran{\c c}ois and Symonds, Graham},
  title   = {Free and Forced Components of Shoaling Long Waves in the Absence of Short-Wave Breaking},
  journal = {Journal of Physical Oceanography},
  year    = {2021},
  volume  = {51},
  number  = {5},
  pages   = {1465--1487},
  doi     = {10.1175/JPO-D-20-0214.1}
}

@article{Contardo2023,
  author  = {Contardo, Stephanie and Lowe, Ryan J. and Dufois, Fran{\c c}ois and Hansen, Jeff E. and Buckley, M. and Symonds, Graham},
  title   = {Free Long-Wave Transformation in the Nearshore Zone through Partial Reflections},
  journal = {Journal of Physical Oceanography},
  year    = {2023},
  volume  = {53},
  number  = {3},
  pages   = {661--681},
  doi     = {10.1175/JPO-D-22-0109.1}
}

@article{Contardo2025,
  author  = {Contardo, Stephanie and Lowe, Ryan J. and Dufois, Fran{\c c}ois and Hansen, Jeff E. and Symonds, Graham},
  title   = {Free Long Wave Generation: Breakpoint Forcing Versus Bound Wave Release},
  journal = {Journal of Geophysical Research: Oceans},
  year    = {2025},
  volume  = {130},
  number  = {7},
  pages   = {e2025JC022377},
  doi     = {10.1029/2025JC022377}
}

@article{ContardoSymonds2013,
  author  = {Contardo, Stephanie and Symonds, Graham},
  title   = {Infragravity response to variable wave forcing in the nearshore},
  journal = {Journal of Geophysical Research: Oceans},
  year    = {2013},
  volume  = {118},
  number  = {12},
  pages   = {7095--7106},
  doi     = {10.1002/2013JC009430}
}

@article{deBakker2016IG,
  author  = {{de Bakker}, Anouk T. M. and Tissier, Marion F. S. and Ruessink, B. G.},
  title   = {Beach steepness effects on nonlinear infragravity-wave interactions: A numerical study},
  journal = {Journal of Geophysical Research: Oceans},
  year    = {2016},
  volume  = {121},
  number  = {1},
  pages   = {554--570},
  doi     = {10.1002/2015JC011268}
}

@article{Dong2009,
  author  = {Dong, Guohai and Ma, Xiaozhou and Perlin, Marc and Ma, Yuxiang and Yu, Bo and Wang, Gang},
  title   = {Experimental study of long wave generation on sloping bottoms},
  journal = {Coastal Engineering},
  year    = {2009},
  volume  = {56},
  number  = {1},
  pages   = {82--89},
  doi     = {10.1016/j.coastaleng.2008.10.002}
}

@article{Guedes2013,
  author  = {Guedes, Rafael M. C. and Bryan, Karin R. and Coco, Giovanni},
  title   = {Observations of wave energy fluxes and swash motions on a low-sloping, dissipative beach},
  journal = {Journal of Geophysical Research: Oceans},
  year    = {2013},
  volume  = {118},
  number  = {7},
  pages   = {3651--3669},
  doi     = {10.1002/jgrc.20267}
}

@article{Guerin2019,
  author  = {Gu{\'e}rin, Thomas and {de Bakker}, Anouk and Bertin, Xavier},
  title   = {On the Bound Wave Phase Lag},
  journal = {Fluids},
  year    = {2019},
  volume  = {4},
  number  = {3},
  pages   = {152},
  doi     = {10.3390/fluids4030152}
}

@article{Hasselmann1962,
  author  = {Hasselmann, Klaus},
  title   = {On the non-linear energy transfer in a gravity-wave spectrum. Part 1. General theory},
  journal = {Journal of Fluid Mechanics},
  year    = {1962},
  volume  = {12},
  number  = {4},
  pages   = {481--500},
  doi     = {10.1017/S0022112062000373}
}

@article{Janssen2003,
  author  = {Janssen, T. T. and Battjes, J. A. and {van Dongeren}, A. R.},
  title   = {Long waves induced by short-wave groups over a sloping bottom},
  journal = {Journal of Geophysical Research: Oceans},
  year    = {2003},
  volume  = {108},
  number  = {C8},
  pages   = {3252},
  doi     = {10.1029/2002JC001515}
}

@incollection{Koren1993,
  author    = {Koren, Barry},
  title     = {A Robust Upwind Discretization Method for Advection, Diffusion and Source Terms},
  booktitle = {Numerical Methods for Advection-Diffusion Problems},
  editor    = {Vreugdenhil, C. B. and Koren, B.},
  series    = {Notes on Numerical Fluid Mechanics},
  volume    = {45},
  pages     = {117--138},
  publisher = {Vieweg},
  address   = {Braunschweig/Wiesbaden},
  year      = {1993}
}

@article{Liao2021,
  author  = {Liao, Zhiling and Li, Shaowu and Liu, Ye and Zou, Qingping},
  title   = {An Analytical Spectral Model for Infragravity Waves over Topography in Intermediate and Shallow Water under Nonbreaking Conditions},
  journal = {Journal of Physical Oceanography},
  year    = {2021},
  volume  = {51},
  number  = {9},
  pages   = {2749--2765},
  doi     = {10.1175/JPO-D-20-0164.1}
}

@article{Liao2023,
  author  = {Liao, Zhiling and Zou, Qingping and Liu, Ye and Contardo, Stephanie and Li, Shaowu},
  title   = {Unified analytical solution for group-induced infragravity waves based on {Green's} function},
  journal = {Journal of Fluid Mechanics},
  year    = {2023},
  volume  = {967},
  pages   = {A37},
  doi     = {10.1017/jfm.2023.475}
}

@article{Liu2023,
  author  = {Liu, Ye and Yao, Yu and Liao, Zhiling and Li, Shaowu and Zhang, Chi and Zou, Qingping},
  title   = {Fully nonlinear investigation on energy transfer between long waves and short-wave groups over a reef},
  journal = {Coastal Engineering},
  year    = {2023},
  volume  = {179},
  pages   = {104240},
  doi     = {10.1016/j.coastaleng.2022.104240}
}

@article{LonguetHiggins1962,
  author  = {Longuet-Higgins, M. S. and Stewart, R. W.},
  title   = {Radiation stress and mass transport in gravity waves, with application to `surf beats'},
  journal = {Journal of Fluid Mechanics},
  year    = {1962},
  volume  = {13},
  number  = {4},
  pages   = {481--504},
  doi     = {10.1017/S0022112062000877}
}

@article{Marchesiello2026,
  author  = {Marchesiello, Patrick and Klotz, Adrien and Pezerat, Marc and Treillou, Simon and Almar, Rafael},
  title   = {Flashrip dynamics in the surfzone: Contrasting wave- and group-resolving models},
  journal = {Ocean Modelling},
  year    = {2026},
  volume  = {202},
  pages   = {102750},
  doi     = {10.1016/j.ocemod.2026.102750}
}

@article{Masselink2019,
  author  = {Masselink, Gerd and Tuck, Megan and McCall, Robert and {van Dongeren}, Ap and Ford, Murray and Kench, Paul},
  title   = {Physical and Numerical Modeling of Infragravity Wave Generation and Transformation on Coral Reef Platforms},
  journal = {Journal of Geophysical Research: Oceans},
  year    = {2019},
  volume  = {124},
  number  = {3},
  pages   = {1410--1433},
  doi     = {10.1029/2018JC014411}
}

@article{Matsuba2021,
  author  = {Matsuba, Yoshinao and Shimozono, Takenori and Tajima, Yoshimitsu},
  title   = {Tidal modulation of infragravity wave dynamics on a reflective barred beach},
  journal = {Estuarine, Coastal and Shelf Science},
  year    = {2021},
  volume  = {261},
  pages   = {107562},
  doi     = {10.1016/j.ecss.2021.107562}
}

@article{McCall2010,
  author  = {McCall, R. T. and {van Thiel de Vries}, J. S. M. and Plant, N. G. and {van Dongeren}, A. R. and Roelvink, J. A. and Thompson, D. M. and Reniers, A. J. H. M.},
  title   = {Two-dimensional time dependent hurricane overwash and erosion modeling at Santa Rosa Island},
  journal = {Coastal Engineering},
  year    = {2010},
  volume  = {57},
  number  = {7},
  pages   = {668--683},
  doi     = {10.1016/j.coastaleng.2010.02.006}
}

@article{MeiBenmoussa1984,
  author  = {Mei, Chiang C. and Benmoussa, Chakib},
  title   = {Long waves induced by short-wave groups over an uneven bottom},
  journal = {Journal of Fluid Mechanics},
  year    = {1984},
  volume  = {139},
  pages   = {219--235},
  doi     = {10.1017/S0022112084000331}
}

@article{MihamiRoeber2026,
  author  = {Mihami, Fatima-Zahra and Roeber, Volker},
  title   = {Exploring a conservative staggered scheme for Boussinesq-type equations: Insights into numerical diffusion, dispersion, and wave-breaking},
  journal = {Coastal Engineering},
  year    = {2026},
  volume  = {204},
  pages   = {104880},
  doi     = {10.1016/j.coastaleng.2025.104880}
}

@inproceedings{Midouni2025CD,
  author    = {Midouni, Hedi and Monbaliu, Jaak},
  title     = {Exploring Infragravity Wave Generation Mechanisms Through the Vortex Force Formalism},
  booktitle = {Coastal Dynamics 2025},
  editor    = {Coelho, Carlos and Hallin, Caroline and Sancho, Francisco and Silva, Paulo A.},
  series    = {Coastal Research Library},
  volume    = {42},
  pages     = {77--83},
  publisher = {Springer},
  address   = {Cham},
  year      = {2026},
  doi       = {10.1007/978-3-032-15477-4_13}
}

@article{Miles1974,
  author  = {Miles, J. W.},
  title   = {Harbor Seiching},
  journal = {Annual Review of Fluid Mechanics},
  year    = {1974},
  volume  = {6},
  pages   = {17--33},
  doi     = {10.1146/annurev.fl.06.010174.000313}
}

@article{Moura2018,
  author  = {Moura, T. and Baldock, T. E.},
  title   = {New Evidence of Breakpoint Forced Long Waves: Laboratory, Numerical, and Field Observations},
  journal = {Journal of Geophysical Research: Oceans},
  year    = {2018},
  volume  = {123},
  number  = {4},
  pages   = {2716--2730},
  doi     = {10.1002/2017JC013621}
}

@article{Munk1949,
  author  = {Munk, Walter H.},
  title   = {Surf beats},
  journal = {Transactions, American Geophysical Union},
  year    = {1949},
  volume  = {30},
  number  = {6},
  pages   = {849--854},
  doi     = {10.1029/TR030i006p00849}
}

@inproceedings{Nielsen2017,
  author    = {Nielsen, Peter},
  title     = {Surf beat ``shoaling''},
  booktitle = {Coastal Dynamics 2017},
  address   = {Helsing{\o}r, Denmark},
  pages     = {443--450},
  year      = {2017}
}

@article{NielsenBaldock2010,
  author  = {Nielsen, Peter and Baldock, Tom E.},
  title   = {{\reflectbox{N}}-Shaped surf beat understood in terms of transient forced long waves},
  journal = {Coastal Engineering},
  year    = {2010},
  volume  = {57},
  number  = {1},
  pages   = {71--73},
  doi     = {10.1016/j.coastaleng.2009.09.003}
}

@article{Okihiro1993,
  author  = {Okihiro, Michele and Guza, R. T. and Seymour, R. J.},
  title   = {Excitation of seiche observed in a small harbor},
  journal = {Journal of Geophysical Research: Oceans},
  year    = {1993},
  volume  = {98},
  number  = {C10},
  pages   = {18201--18211},
  doi     = {10.1029/93JC01760}
}

@article{Olabarrieta2023,
  author  = {Olabarrieta, Maitane and Warner, John C. and Hegermiller, Christie A.},
  title   = {Development and Application of an Infragravity Wave ({InWave}) Driver to Simulate Nearshore Processes},
  journal = {Journal of Advances in Modeling Earth Systems},
  year    = {2023},
  volume  = {15},
  number  = {6},
  pages   = {e2022MS003205},
  doi     = {10.1029/2022MS003205}
}

@article{Pomeroy2012,
  author  = {Pomeroy, Andrew and Lowe, Ryan and Symonds, Graham and {van Dongeren}, Ap and Moore, Christine},
  title   = {The dynamics of infragravity wave transformation over a fringing reef},
  journal = {Journal of Geophysical Research: Oceans},
  year    = {2012},
  volume  = {117},
  number  = {C11},
  pages   = {C11022},
  doi     = {10.1029/2012JC008310}
}

@incollection{Rabinovich2009,
  author    = {Rabinovich, Alexander B.},
  title     = {Seiches and Harbor Oscillations},
  booktitle = {Handbook of Coastal and Ocean Engineering},
  editor    = {Kim, Young C.},
  publisher = {World Scientific},
  year      = {2009},
  pages     = {193--236},
  doi       = {10.1142/9789812819307_0009}
}

@article{Reyns2023,
  author  = {Reyns, Johan and McCall, Robert and Ranasinghe, Roshanka and {van Dongeren}, Ap and Roelvink, Dano},
  title   = {Modelling wave group-scale hydrodynamics on orthogonal unstructured meshes},
  journal = {Environmental Modelling \& Software},
  year    = {2023},
  volume  = {162},
  pages   = {105655},
  doi     = {10.1016/j.envsoft.2023.105655}
}

@article{Rijnsdorp2022,
  author  = {Rijnsdorp, Dirk P. and Smit, Pieter B. and Guza, R. T.},
  title   = {A nonlinear, non-dispersive energy balance for surfzone waves: Infragravity wave dynamics on a sloping beach},
  journal = {Journal of Fluid Mechanics},
  year    = {2022},
  volume  = {944},
  pages   = {A45},
  doi     = {10.1017/jfm.2022.512}
}

@article{RoeberBricker2015,
  author  = {Roeber, Volker and Bricker, Jeremy D.},
  title   = {Destructive tsunami-like wave generated by surf beat over a coral reef during Typhoon Haiyan},
  journal = {Nature Communications},
  year    = {2015},
  volume  = {6},
  pages   = {7854},
  doi     = {10.1038/ncomms8854}
}

@article{Roelvink2009,
  author  = {Roelvink, Dano and Reniers, Ad and {van Dongeren}, Ap and {van Thiel de Vries}, Jaap and McCall, Robert and Lescinski, Jamie},
  title   = {Modelling storm impacts on beaches, dunes and barrier islands},
  journal = {Coastal Engineering},
  year    = {2009},
  volume  = {56},
  number  = {11--12},
  pages   = {1133--1152},
  doi     = {10.1016/j.coastaleng.2009.08.006}
}

@article{Ruessink1998,
  author  = {Ruessink, B. G.},
  title   = {Bound and free infragravity waves in the nearshore zone under breaking and nonbreaking conditions},
  journal = {Journal of Geophysical Research: Oceans},
  year    = {1998},
  volume  = {103},
  number  = {C6},
  pages   = {12795--12805},
  doi     = {10.1029/98JC00893}
}

@article{Russell1993,
  author  = {Russell, P. E.},
  title   = {Mechanisms for beach erosion during storms},
  journal = {Continental Shelf Research},
  year    = {1993},
  volume  = {13},
  number  = {11},
  pages   = {1243--1265},
  doi     = {10.1016/0278-4343(93)90051-X}
}

@article{Schaeffer1993,
  author  = {Sch{\"a}ffer, H. A.},
  title   = {Infragravity waves induced by short-wave groups},
  journal = {Journal of Fluid Mechanics},
  year    = {1993},
  volume  = {247},
  pages   = {551--588},
  doi     = {10.1017/S0022112093000564}
}

@article{Smith2006,
  author  = {Smith, Jerome A.},
  title   = {Wave-Current Interactions in Finite Depth},
  journal = {Journal of Physical Oceanography},
  year    = {2006},
  volume  = {36},
  number  = {7},
  pages   = {1403--1419},
  doi     = {10.1175/JPO2911.1}
}

@article{Smit2013,
  title = {Depth-induced wave breaking in a non-hydrostatic,  near-shore wave model},
  volume = {76},
  ISSN = {0378-3839},
  url = {http://dx.doi.org/10.1016/j.coastaleng.2013.01.008},
  DOI = {10.1016/j.coastaleng.2013.01.008},
  journal = {Coastal Engineering},
  publisher = {Elsevier BV},
  author = {Smit,  Pieter and Zijlema,  Marcel and Stelling,  Guus},
  year = {2013},
  month = {June},
  pages = {1–16}
}

@article{StellingDuinmeijer2003,
  author  = {Stelling, G. S. and Duinmeijer, S. P. A.},
  title   = {A staggered conservative scheme for every Froude number in rapidly varied shallow water flows},
  journal = {International Journal for Numerical Methods in Fluids},
  year    = {2003},
  volume  = {43},
  number  = {12},
  pages   = {1329--1354},
  doi     = {10.1002/fld.537}
}

@article{Stockdon2006,
  author  = {Stockdon, H. F. and Holman, R. A. and Howd, P. A. and Sallenger, A. H. Jr.},
  title   = {Empirical parameterization of setup, swash, and runup},
  journal = {Coastal Engineering},
  year    = {2006},
  volume  = {53},
  number  = {7},
  pages   = {573--588},
  doi     = {10.1016/j.coastaleng.2005.12.005}
}

@article{Symonds1982,
  author  = {Symonds, G. and Huntley, D. A. and Bowen, A. J.},
  title   = {Two-dimensional surf beat: Long wave generation by a time-varying breakpoint},
  journal = {Journal of Geophysical Research},
  year    = {1982},
  volume  = {87},
  number  = {C1},
  pages   = {492--498},
  doi     = {10.1029/JC087iC01p00492}
}

@article{Tucker1950,
  author  = {Tucker, M. J.},
  title   = {Surf beats: Sea waves of 1 to 5 min. period},
  journal = {Proceedings of the Royal Society of London. Series A, Mathematical and Physical Sciences},
  year    = {1950},
  volume  = {202},
  number  = {1071},
  pages   = {565--573},
  doi     = {10.1098/rspa.1950.0120}
}

@article{Uchiyama2010,
  author  = {Uchiyama, Yusuke and McWilliams, James C. and Shchepetkin, Alexander F.},
  title   = {Wave-current interaction in an oceanic circulation model with a vortex-force formalism: Application to the surf zone},
  journal = {Ocean Modelling},
  year    = {2010},
  volume  = {34},
  number  = {1--2},
  pages   = {16--35},
  doi     = {10.1016/j.ocemod.2010.04.002}
}

@article{vanDongeren2003,
  author  = {{van Dongeren}, A. R. and Reniers, A. J. H. M. and Battjes, J. A. and Svendsen, I. A.},
  title   = {Numerical modeling of infragravity wave response during {DELILAH}},
  journal = {Journal of Geophysical Research: Oceans},
  year    = {2003},
  volume  = {108},
  number  = {C9},
  pages   = {3288},
  doi     = {10.1029/2002JC001332}
}

@article{vanDongeren2007,
  author  = {{van Dongeren}, A. and Battjes, J. and Janssen, T. and {van Noorloos}, J. and Steenhauer, K. and Steenbergen, G. and Reniers, A.},
  title   = {Shoaling and shoreline dissipation of low-frequency waves},
  journal = {Journal of Geophysical Research: Oceans},
  year    = {2007},
  volume  = {112},
  number  = {C2},
  pages   = {C02011},
  doi     = {10.1029/2006JC003701}
}

@article{Vasarmidis2024,
  title = {A study of the non-linear properties and wave generation of the multi-layer non-hydrostatic wave model SWASH},
  volume = {302},
  ISSN = {0029-8018},
  url = {http://dx.doi.org/10.1016/j.oceaneng.2024.117633},
  DOI = {10.1016/j.oceaneng.2024.117633},
  journal = {Ocean Engineering},
  publisher = {Elsevier BV},
  author = {Vasarmidis,  Panagiotis and Klonaris,  Georgios and Zijlema,  Marcel and Stratigaki,  Vasiliki and Troch,  Peter},
  year = {2024},
  month = {June},
  pages = {117633}
}

@article{Zhang2020,
  author  = {Zhang, Q. and Toorman, E. A. and Monbaliu, J.},
  title   = {Shoaling of bound infragravity waves on plane slopes for bichromatic wave conditions},
  journal = {Coastal Engineering},
  year    = {2020},
  volume  = {158},
  pages   = {103684},
  doi     = {10.1016/j.coastaleng.2020.103684}
}

@article{Zou2011,
  author  = {Zou, Qingping},
  title   = {Generation, Transformation, and Scattering of Long Waves Induced by a Short-Wave Group over Finite Topography},
  journal = {Journal of Physical Oceanography},
  year    = {2011},
  volume  = {41},
  number  = {10},
  pages   = {1842--1859},
  doi     = {10.1175/2011JPO4511.1}
}

@article{Zijlema2011,
  title = {SWASH: An operational public domain code for simulating wave fields and rapidly varied flows in coastal waters},
  volume = {58},
  ISSN = {0378-3839},
  url = {http://dx.doi.org/10.1016/j.coastaleng.2011.05.015},
  DOI = {10.1016/j.coastaleng.2011.05.015},
  number = {10},
  journal = {Coastal Engineering},
  publisher = {Elsevier BV},
  author = {Zijlema,  Marcel and Stelling,  Guus and Smit,  Pieter},
  year = {2011},
  month = Oct,
  pages = {992–1012}
}

@book{Phillips1977,
  author    = {Phillips, O. M.},
  title     = {The Dynamics of the Upper Ocean},
  edition   = {2},
  publisher = {Cambridge University Press},
  address   = {Cambridge},
  year      = {1977}
}

@article{ShyuPhillips1990,
  author  = {Shyu, Jinn-Hwa and Phillips, O. M.},
  title   = {The blockage of gravity and capillary waves by longer waves and currents},
  journal = {Journal of Fluid Mechanics},
  volume  = {217},
  pages   = {115--141},
  year    = {1990},
  doi     = {10.1017/S0022112090000659}
}

\end{document}